\documentclass[twocolumn,letterpaper,amsmath,amssymb,floatfix,aps,superscriptaddress]{revtex4} 

\usepackage{graphicx}
\usepackage{dcolumn}
\usepackage{bm}
\usepackage{epsfig,color}

\newcommand{\Av}[1]{{\mathbf #1}}

\def\ln{{\operatorname{ln}}}

\def\rmd{{\mathrm{d}}}

\def\rme{{\mathrm{e}}}

\begin{document}

\title{Counterion-mediated weak and strong coupling electrostatic interaction between like-charged cylindrical dielectrics}

\author{Matej Kandu\v c}
\affiliation{Department of Theoretical Physics,
J. Stefan Institute, SI-1000 Ljubljana, Slovenia}

\author{Ali Naji}
\affiliation{Department of Physics and Astronomy, University of Sheffield, Sheffield S3 7RH, United Kingdom}

\author{Rudolf Podgornik}
\affiliation{Department of Theoretical Physics,
J. Stefan Institute, SI-1000 Ljubljana, Slovenia}
\affiliation{Institute of Biophysics, Medical Faculty and Department of Physics, Faculty of Mathematics and Physics,
University of Ljubljana, SI-1000 Ljubljana, Slovenia}

\begin{abstract}
We examine the effective  counterion-mediated electrostatic interaction between two like-charged dielectric cylinders immersed in a continuous dielectric medium containing neutralizing mobile counterions. 
We focus on the effects of image charges induced as a result of the dielectric mismatch between the cylindrical cores and the surrounding dielectric medium 
and investigate the counterion-mediated electrostatic interaction between the cylinders in both limits of weak and strong electrostatic couplings 
(corresponding, e.g., to systems with monovalent and multivalent counterions, respectively). The results are compared with extensive Monte-Carlo 
simulations exhibiting good agreement with the limiting weak and strong coupling results in their respective regime of validity. 
\end{abstract}

\maketitle

\section{Introduction}

Electrostatic interactions of charged macromolecules and colloids are often governed by small oppositely charged ions (counterions) that maintain global electroneutrality. These counterion-mediated interactions play a fundamental role in classical charged (Coulomb) fluids that are abundant in biological and soft matter context \cite{holm,Andelman} and include many charged macromolecules (such as nucleic acids DNA and RNA, actin filaments, microtubules and globular proteins), affecting their functional, structural and dynamical behavior.  
In spite of the importance of electrostatic interactions, there is no general method that would allow for an accurate prediction of electrostatic effects  in all regions of the 
parameter space, defined by the surface charge density of macroions, charge valency of counterions, dielectric mismatches between the often hydrophobic core of the macromolecule and the surrounding aqueous medium, etc. 
 Often the electrostatic interactions are treated on the 
Poisson-Boltzmann (PB) level leading to 
 effective interactions which turn out to be always repulsive between like-charged macromolecules. 
Conceptually, the PB approach corresponds to a mean-field treatment of electrostatic interactions and
is asymptotically valid  only for sufficiently large separations between macromolecules, low enough surface charge densities and low counterion valency \cite{holm}. It characterizes a 
situation where electrostatic fluctuations and correlations due to the counterions are negligible. There are other regions in 
the parameter space of charged macromolecules where one expects the mean-field 
framework to break down leading to the emergence of a completely different non-PB-type physics. 
A notorious example is the phenomenon of like-charge attraction, which emerges between highly charged macroions or in the presence of 
high valency counterions and has been at the focus of both experimental \cite{Bloom,Strey98,Tang,Tang03,rau-1,rau-2, besteman, Nord} and theoretical investigations 
in recent years (see Refs. \cite{Rouzina,holm,hoda,shklovskii,Levin,rudi-rmp,Netz,Naji,Messina,Lau01, kjellander} for an extensive reference list).

It appears to us that among the most important recent advances in this field has been the systematization of  non-PB effects based on the notions of {\em weak coupling} (WC) 
and {\em strong coupling} (SC) approximations. These terms refer to the strength of electrostatic coupling in the system and may be understood conceptually in terms of the 
electrostatic interactions of mobile counterions  with fixed external charges (macroions) in the system when compared with direct electrostatic interactions between the counterions themselves. This latter contribution 
may be characterized in terms of  the {\em Bjerrum length},
\begin{equation}
\ell_{\mathrm{B}}=e_0^2/(4\pi\varepsilon\varepsilon_0 k_{\mathrm B}T),
\end{equation}
which corresponds to the separation at which two unit charges, $e_0$, interact with thermal energy $k_{\mathrm B}T$ in a medium of dielectric constant $\varepsilon$  (in water and at room temperature, the value is $\ell_{\mathrm{B}}\approx 0.7$ nm). 
If the charge of the counterions is $+qe_0$ then the Bjerrum length scales as $q^2 \ell_{\mathrm{B}}$. The interaction of counterions with macroion charges (of surface charge density $-\sigma_s$)  can be characterized in terms
of the so-called {\em Gouy-Chapman length}, 
\begin{equation}
\mu=e_0/(2\pi q\ell_{\mathrm{B}}\sigma_s ),  
\end{equation}
which gives the separation at which the counterion-surface interaction energy equals $k_{\mathrm B}T$. The ratio of these two fundamental length scales introduces a dimensionless
parameter 
\begin{equation}
\Xi=q^2 \ell_{\mathrm{B}} /\mu,
\label{eq:Xicouple}
\end{equation}
which is known as the (Netz-Moreira) electrostatic coupling parameter  $\Xi$ \cite{Netz} and quantifies the strength of electrostatic coupling in the system. This parameter is closely
related to the plasma parameter of ionic systems \cite{kalman} and may be written also in terms of the typical lateral spacing, $a_\bot$, between counterions in the proximity of a charged surface, i.e.,
$a_\bot/\mu \sim \sqrt{\Xi}$ as set by the local electroneutrality condition $a_\bot^2\sim qe_0/\sigma_s $. 

It then follows that in the weak coupling regime, defined by ${\Xi}\ll 1$, the width of the Gouy-Chapman layer $\mu$ is much larger than the separation between two neighbouring counterions in the counterion layer and thus it behaves basically as a three-dimensional gas. Each counterion in this case interacts with many others and the collective mean-field approach of the Poisson-Boltzmann type is thoroughly justified. On the other hand, in the strong coupling regime, defined by ${\Xi}\gg 1$, the mean distance between counterions, $a_\bot$, exceeds the layer width and thus the 
counterion layer behaves essentially as a two-dimensional layer \cite{Netz}. In this case the mean-field approach breaks down as each counterion is isolated laterally in a relatively large correlation 
hole of size $a_\bot$. In fact, as each counterion can move almost independently from the others along the direction perpendicular to the charged surface, the properties of the system
are dominated by single-particle contributions on the leading order, which is in stark contrast with the collective mean-field picture and emerges as a direct consequence of 
strong electrostatic correlations in the system. The two dychotomous limits are characterised by a low (high) valency of the counterions, a small (large) value of the surface charge density
and/or large (small) medium dielectric constant.

Conceptually the study of the SC regime has been pioneered in several recent works \cite{Rouzina,shklovskii,Levin,Netz,rudi-rmp,Naji,hoda, Lau01} using various analytical methods. 
It was shown \cite{Netz} that both the WC and the SC limits may be described analytically as two exact {\em limiting laws} from a single  general theory for classical 
Coulomb fluids: while the PB theory follows in the limit of $\Xi\rightarrow 0$, a limiting single-particle SC theory follows in the limit $\Xi\rightarrow \infty$, 
which thus  allow for an exact statistical mechanical treatment of charged systems at two opposed limiting conditions. 
The parameter space in between can be analyzed by approximate methods \cite{Forsman04,intermediate_regime,Weeks}, being accessible effectively only {\em via} computer simulations
\cite{hoda,Naji,Netz,asim,Weeks,Jho1,original_sims,Forsman04,trulsson,jho-prl, NajiArnold,Naji_CCT}. 
Exact solutions  for the whole range of coupling parameters are available only in one
dimension \cite{exact}. The WC-SC paradigm has been tested extensively \cite{hoda,Naji,Netz,NajiArnold,asim,Forsman04,intermediate_regime,Weeks,Jho1,jho-prl,Naji_CCT,exact} and was found to describe computer simulations quantitatively correctly in the respective regimes of validity, thus providing a unifying view of the behavior of Coulomb fluids. 
The main facets of these results are retained even when the model system is generalized in order to include more realistic features describing the bathing solution or the nature of the fixed or mobile charges in the system. Though, for instance, multipolar charge distribution of mobile counterions \cite{MULTIPOLES} or statistically disordered distribution of fixed charges \cite{EPLRUDI-ALI} unavoidably introduce novel features in the counterion-mediated electrostatic interaction, the application of the same general philosophy embodied in the weak and strong coupling limits remains sound and valid. The only case where it needs to be amended in an essential manner is when the bathing solution contains a mixture of univalent as well as polyvalent salts, which incidentally are also the most common experimental conditions. In that case a more sophisticated mixed weak-strong coupling analysis is in order leading to qualitatively different results \cite{olli, SC-DHMATEJ}.

Though originally formulated in the context of planar macroions, these advances have transpired also in the DNA-like models of polyelectrolytes which deal with electrostatic interactions in the context of macroions with cylindrical or indeed helical fixed charge distribution. Indeed this particular variant of counterion-mediated interactions has a venerable history \cite{Oosawa, lifson,Alfrey51,Fuoss51,Manning97}. 
During the last two decades several distinct analytical approaches aimed in different directions improved the classical PB results for simple DNA-like models and revealed the importance of correlation effects as well as several other factors including the discrete or helical charge distributions, chain flexibility, finite molecular size, and dielectric inhomogeneities \cite{Lyub95, Gron97,Deserno03,NajiArnold, Barrat, Podgornik98,Ha, Stevens90, Diehl99,Korny, Arenzon99,Shklovs99, Golestan99, gavryushov1, gavryushov2, NajiNetzEPJE, kanduc-helix,Trizac,Muthu04,Liu_Muthu02,Manghi,Henle04,hansen, arnoldholm,Naji_CCT,cherstvy,lee,holm_images}.
Guided by these developments we set ourselves the goal of systematically analyzing the interactions between cylindrical macroions mediated by mobile counterions in the presence of dielectric inhomogeneities.

Contrary to the case of planar macroions which can be completely characterized by a single dimensionless coupling parameter, cylindrical macroions require in general two dimensionless coupling parameters that consistently describe the range of validity of the strong and weak coupling approximations. The existence of two coupling parameters is due to the simple fact that,  if compared to the planar case, the cylinder has a finite radius that needs to enter the fundamental description of the problem.
This other dimensionless parameter brought fourth by cylindrical geometry of the macroion
is nothing but the so-called {\em Manning parameter} \cite{Manning69}. For a cylinder of radius $a$ and linear charge density $\lambda$, it is given by 
\begin{equation}
\xi=q \ell_{\mathrm{B}} \frac{\lambda}{e_0}=\frac{a}{\mu}. 
\label{eq:xi_M}
\end{equation}
The Manning parameter thus represents the dimensionless linear charge density or, 
on the other hand,  also  the rescaled radius of the cylindrical charge distribution. 
Note that the two parameters (i.e., $\Xi$ and $\xi$) in the cylindrical geometry are independent and can be set separately.
For double-stranded DNA ($\lambda\approx 6 \,e_0/{\mathrm{nm}}$), the Manning parameter and the (Netz-Moreira) electrostatic coupling parameter are given by $\xi\simeq 4.1\,q$ and $\Xi\simeq 2.8\,q^3$ (at room temperature in water) in terms of the counterion valency $q$. 

In this paper we will analyze systematically the interactions between two cylindrical macroions characterized by a fixed surface charge density as well as a finite dielectric jump between the external dielectric background (corresponding to a continuum solvent) 
and the hydrophobic cores of cylindrical macroions. Analytically derived results in both limits of strong and weak 
coupling will be compared  directly with Monte-Carlo simulation results.
The organization of this paper is as follows: We first introduce our model and study the image-charge effects within the 
problem of a single charged cylinder with neutralizing counterions in Section \ref{sec:one_cyl}. We then focus on the 
interaction between two identical charged dielectric cylinders within the WC and the SC theory as well as MC simulations  
in Section \ref{sec:two_cylinder}. The results  in the presence and absence of dielectric discontinuity at the cylindrical 
boundaries are analyzed in detail in order to bring up the effects due to the image charges in the two-cylinder geometry.

\section{One charged cylinder}
\label{sec:one_cyl}

Let us focus first on the problem of a single infinitely long uniformly charged cylindrical macroion of radius $a$. The charge of the cylinder is assumed to be distributed uniformly  on its  surface according to  the charge distribution function 
\begin{equation}
\sigma(\Av r)=-\frac{\lambda}{2\pi a}\,\delta(\rho-a), 
\label{n0}
\end{equation}
with $\lambda$ being the absolute {\em linear} charge density and $\rho$ the radial coordinate from the cylinder axis. The cylinder is arbitrarily chosen to be negatively charged and its electrical charge is thus neutralized by positively charged mobile counterions of valency $+q$, which are present in the region $\rho>a$ only. 
\begin{figure}[h]
\centerline{\psfig{figure=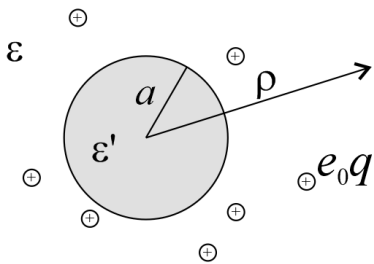,width=5cm}}
\caption{Schematic top view of a single charged dielectric cylinder of radius $a$ and dielectric constant $\varepsilon'$, along with mobile neutralizing counterions of charge valency $+q$ dispersed in a continuum solvent of dielectric constant $\varepsilon$. 
The system is confined coaxially in a cylindrical box of radius $a_{\mathrm{out}}$ (not shown).}
\label{fig:model_1cyln}
\end{figure}

In general, the interior of the cylindrical macromolecule can be characterized by a different dielectric constant $\varepsilon'$ than the surrounding continuum solvent medium, $\varepsilon$. This is certainly the case for DNA that has a hydrophobic inner core composed mostly of stacked nitrogen bases that have a vastly different dielectric response from an aqueous solution \cite{rudi-rmp}. Other charged polymers that usually do not share structural features with DNA can nevertheless also posses hydrophobic inner cores with a local dielectric response that differs from the one of the solvent. The hydrophobic apolar core of the macromolecular backbone would then have a static dielectric constant $\varepsilon'\simeq 2$ (hydrocarbon), while the surrounding continuum dielectric medium which is usually water, a polar associated liquid, would be characterized by $\varepsilon\simeq 80$. One can thus introduce the dielectric discontinuity parameter 
\begin{equation}
\Delta=\frac{\varepsilon-\varepsilon'}{\varepsilon+\varepsilon'},
\label{eq:jump}
\end{equation}
which measures the relative dielectric mismatch at the interface of the two materials. 
In a dielectrically homogeneous system, we have $\Delta=0$ and  no image
charge effects are present, while, for instance, in water-hydrocarbon
systems, one has $\Delta=0.95$, which suggests strong image charge
effects (note that $|\Delta|\leq 1$ and the largest value for $\Delta$ is 
obtained when one medium is ideally polarizable, i.e., is an ideal metal).
Therefore, a single cylinder can be described by three different dimensionless parameters, namely, the electrostatic coupling parameter $\Xi$,  the  Manning parameter $\xi$ and the dielectric discontinuity parameter $\Delta$ as defined in Eqs.~(\ref{eq:Xicouple}), (\ref{eq:xi_M}) and (\ref{eq:jump}), respectively. 

The presence of a dielectric inhomogeneity across the boundary of the cylinder, see Fig.~\ref{fig:model_1cyln}, influences the electrostatic potential that can be thus described by the Green's function connecting two points $\Av r, \Av r'$ outside the cylinder as
\begin{equation}
u(\Av r, \Av r')=u_0(\Av r, \Av r')+u_\textrm{im}(\Av r, \Av r'),
\label{eq:u_0u_im}
\end{equation}
where $u_0$ is the direct  standard Coulomb kernel in the absence of  dielectric inhomogeneities,
\begin{equation}
u_0(\Av r, \Av r')=\frac{1}{4\pi\varepsilon\varepsilon_0\vert\Av r-\Av r'\vert},
\end{equation}
and $u_{\rm im}$ is the  ``image correction" due to the dielectric jump  just as in the case of a planar discontinuity. Unfortunately in cylindrical geometry the concept of Kelvin image charges, that plays such a fundamental role for planar dielectric discontinuities, is a bit diluted since the image correction can not be in general formulated in a way that would entail a summation of properly positioned point image charges. We nevertheless consistently refer to the effects of dielectric inhomogeneities in this system as dielectric image effects. If explicitly stated we believe this inconsistency in nomenclature can not be a source of confusion.

Instead of using the concept of point image charges one must in fact explicitly solve the Poisson equation in cylindrical geometry specified by cylindrical coordinates $(\rho, \varphi, z)$ with appropriate boundary conditions at the surface of the cylinder, which requires the normal component of the electric field to fulfill the following relation
\begin{equation}
\varepsilon\varepsilon_0 E_n|_{\rho=a^+}-\varepsilon'\varepsilon_0 E_n|_{\rho=a^-}=\sigma_s,
\label{bc}
\end{equation}
where 
\begin{equation}
\sigma_s=\lambda/(2\pi a),
\end{equation}
is the cylinder surface charge density. Solving the Poisson equation with the proper eigenfunction expansion \cite{smythe} one can express the image part of the Green's function as
\begin{eqnarray}
u_\textrm{im}(\Av r,\Av r')&=&\frac{1}{2\pi^2\varepsilon\varepsilon_0}\sum_{m=0}^\infty\int_0^\infty\rmd k\,
\xi_m(ka)K_m(k\rho')K_m(k\rho)\nonumber\\
&&\times\>\cos(k\Delta z)\,\cos(m\Delta\varphi),
\label{Vim}
\end{eqnarray}
with the following definition
\begin{equation}
\xi_m(x)\equiv\frac{2 (2-\delta_{m0})\Delta I_m(x)}{\frac{1+\Delta}{x I_{m+1}(x)+mI_m(x)}-2\Delta K_m(x)},
\end{equation}
where $\Delta\varphi$ and $\Delta z$ are the angle and height differences between the position vectors $\Av r$ and $\Av r'$, respectively. $I_{m}(x)$ and $K_m(x)$ are the modified Bessel functions of the first and the second kind that enter the eigenfunction expansion \cite{smythe}.

Having derived the appropriate  Green's function for the cylindrical geometry, we now investigate the behavior of counterions in proximity to a charged dielectric cylinder in the two limits specified by the weak and strong coupling approximations that can be treated analytically. The behavior at intermediate coupling strengths will be analyzed by using extensive MC simulations. 

\subsection{Weak coupling limit: Mean-field theory}

As noted before, the WC regime is characterized by a small coupling parameter ${\Xi}\ll 1$, which is adequate for low valency counterions, small surface charge, high temperature or high solvent dielectric constant.   
In the strict limit of $\Xi\rightarrow 0$, ignoring the usually small fluctuations around the mean field configuration, the system is described exactly by the mean-field PB theory \cite{hoda}. Formally the PB equation corresponds to the saddle-point condition imposed on the action field-functional in this limit \cite{podgornik}. For the present system, the PB equation governing the mean electrostatic potential takes the form
\begin{equation}
-\frac{\varepsilon\varepsilon_0}{e_0q}\nabla^2 \phi = \left\{
	 \begin{array}{ll}
	C\, \rme^{-\beta e_0q\phi(\Av r)}&\qquad \rho> a, \\ 
		\\
	0& \qquad \rho<a, 
	\end{array}
	\right. 
	\label{PB}
\end{equation}
where the right-hand side is obviously the number density of the counterions, i.e.
\begin{equation}
n(\Av r) = C\, \rme^{-\beta e_0q\phi(\Av r)}, \qquad \rho> a. 
\end{equation}
Note that due to axial symmetry and translational invariance in the $z$ direction, the problem is reduced to a one-dimensional formulation with $\nabla^2 = {\nabla_{\rho}}^2$ and the solution of the PB equation depends only on the radial coordinate $\rho$. The analytical solution of the PB equation in this case is well known and was first discussed in Refs. \cite{Alfrey51,Fuoss51,lifson} within the so-called {\em cell model} where the cylindrical macroion is confined (coaxially) in an outer cylindrical boundary of radius $a_{\mathrm{out}}$.  The cell model guarantees a finite solution for the counterion density despite the extremely slowly varying (logarithmic) electrostatic potential in two dimensions.

The solution of the PB equation for the counterion density takes different forms depending on whether the Manning parameter is larger or smaller than a threshold value given by $\Lambda \equiv \ln (a_{\mathrm{out}}/a)/[1 + \ln (a_{\mathrm{out}}/a)]$ \cite{Alfrey51,Fuoss51,lifson}. Here we are interested mainly in the situations with sufficiently large Manning parameters $\xi> \Lambda $, where the normalized counterion density can be written as 
\begin{equation}
\tilde n(\rho)= \frac{\alpha^2}{2\pi \xi \rho^2}\sin^{-2}\bigg[\alpha\, \ln \bigg(\frac {\rho}{a} \bigg) +\cot^{-1}\frac{\xi-1}{\alpha}\bigg], 
\label{nPB}
\end{equation}
with $\alpha$ determined from the transcendental equation 
\begin{equation}
\xi   = \frac{1+\alpha^2}{1-\alpha \cot[-\alpha\, \ln (a_{\mathrm{out}}/a)]}. 
\end{equation}
Note that the density profile is renormalized such that we have 
\begin{equation}
2\pi\int_a^{a_{\mathrm{out}}}\tilde n(\rho)\,\rho\rmd\rho=1.
\label{normrho}
\end{equation}
The above density distribution is a monotonically decaying function with very slowly convergent asymptotics \cite{hansen}. It is also independent of the dielectric jump parameter, $\Delta$, due to the axial symmetry which implies that  the electrostatic field vanishes inside the cylinder. The absence of dielectric discontinuity effects is specific to the mean-field limit where no fluctuations are taken into account and can be derived also in the case of planar slabs \cite{kanduc-epje}.

When the Manning parameter is decreased, the system exhibits a continuous counterion condensation transition \cite{Manning69} at a critical Manning parameter $\xi= 1$. The nature of this transition has been analyzed throughly by means of analytical and numerical methods \cite{Naji_CCT}. It was in particular shown that the behavior close to the transition point is described by the mean-field  theory and fluctuation and correlation effects (that will be important eventually at sufficiently large Manning
parameters) play no role at the transition point. Here we shall not consider the behavior of counterions close to
the transition point but do note that since the image-charge effects are absent in the mean-field limit, such effects are expected to have no influence on the counterion condensation transition itself. 

In what follows, we shall also take into account the hard-core repulsion between counterions and the cylinder by assuming that the counterions have a finite radius $R_c$. Within the PB theory, this can not be done explicitly unless full corrections due to excluded-volume repulsions between counterions are taken into account \cite{borukhov,Bhuiyan}. Here we shall consider a simplified version of the ion size effects by taking an effective hard-core radius for the cylinder, i.e., by setting $a\to a+R_c$.  Though approximate on the WC level, this procedure turns out to be exact in the SC limit \cite{NajiNetzEPJE}. 
\begin{figure*}[t]
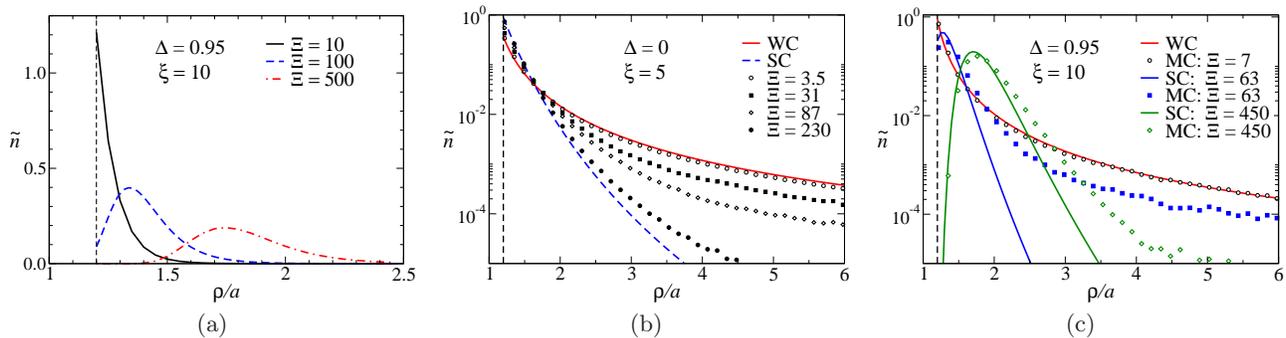
\begin{center}
	\begin{minipage}[b]{0.302\textwidth}\begin{center}
		\includegraphics[width=\textwidth]{rhoSC.eps} (a)
	\end{center}\end{minipage} \hskip0.25cm
	\begin{minipage}[b]{0.302\textwidth}\begin{center}
		\includegraphics[width=\textwidth]{rho5.eps} (b)
	\end{center}\end{minipage} \hskip0.25cm
	\begin{minipage}[b]{0.302\textwidth}\begin{center}
		\includegraphics[width=\textwidth]{rho10im.eps} (c)
	\end{center}\end{minipage} \hskip0.25cm
	\caption{a) Radial counterion number density profile around a single charged cylinder 
	as predicted by the SC theory, Eq.~(\ref{nSC}), for various values of the coupling parameter. Vertical dashed line represents the hard-core exclusion volume $a+R_c$, where counterions radius in all the figures is taken as $R_c=0.2\,a$.
	b) Log-linear plot of the counterion density as obtained from the WC theory (Eq.~(\ref{nPB}), red line), the SC theory (Eq.~(\ref{nSC}), black line) 
	and MC simulations (symbols) in the absence of image-charge effects  ($\Delta=0$). c) Same as (b) but in the presence of a dielectric discontinuity 
	at the cylinder surface with $\Delta=0.95$. The system is considered in a cylindrical confining cell of
	radius  $a_{\mathrm{out}}/a=100$.}
\label{fig:dens}
\end{center}\end{figure*}

\subsection{Strong coupling limit}

In the regime when $\Xi\gg 1$, the mean-field approximation breaks down, and a different kind of approach is needed. We shall focus on the limit of $\Xi\rightarrow \infty$, where the system can be treated via a systematic strong coupling virial expansion \cite{Netz, Naji}  leading to an exact analytical theory in the leading order in $\Xi$. By construction the SC theory is a single particle theory and contains contributions only from single-particle interactions. We do not elaborate on the derivation of the SC theory here as further details can be found in previous publications \cite{Netz,Naji,hoda}.

In this Section, we shall be interested in the counterion density profile around the cylinder 
in the SC limit. The general form of the counterion distribution within the SC theory turns out to be given in terms of a Boltzmann factor containing the interaction energy of an individual counterion with a fixed charged macroion \cite{Netz}.  In a dielectrically inhomogeneous system, such as a charged cylindrical core considered here, the SC counterion density may be written as \cite{kanduc-epje}
\begin{equation}
\tilde n(\Av r)\propto\exp\left[-\beta W_{\rm self}(\Av r)-
\beta W_{0c}(\Av r)\right], 
\label{eq:density_ci}
\end{equation}
where 
\begin{equation}
\beta W_{\rm self}(\Av r)=\frac 12\beta(e_0 q)^2 \,u_\textrm{\rm im}(\Av r,\Av r),
\end{equation}
is the dielectric image {\em self-energy} of a single counterion, i.e., the electrostatic energy of a charged point particle 
in the vicinity of a neutral dielectric cylinder. This energy of course represents the contribution from the interaction of a counterion with its
image charge and thus depends crucially on the value of the dielectric discontinuity at the macroion surface. Furthermore,
\begin{equation}
\beta W_{0c}(\Av r)=\beta e_0q\int u(\Av r,\Av r')\sigma(\Av r')\rmd \Av r'
	=v_0(\Av r)+v_{\rm im}(\Av r), 
\label{eq_W0c}	
\end{equation}
where 
\begin{equation}
v_0(\Av r)=\beta e_0q\int u_0(\Av r,\Av r')\sigma(\Av r')\rmd \Av r'
	= 2\xi\,\ln\,\rho, 
\end{equation}
is the bare electrostatic interaction energy of a single counterion located at a given position $\Av r = (\rho, \varphi, z)$ with the cylinder charge (up to an irrelevant additive constant). 
It goes to zero when the charge density on the macroion surface goes to zero.  The image-dependent part of the interaction energy between a point charge 
and the surface charge distribution on the dielectric cylinder, $v_{\rm im}(\Av r)$, can be shown to be nil due to the axial symmetry, i.e.
\begin{equation}
v_{\rm im}(\Av r)=\beta e_0 q\int u_{\rm im}(\Av r,\Av r')\sigma(\Av r')\rmd \Av r'=0.
\label{vself}
\end{equation}
If the surface charge distribution on the macroion is  non-uniform, such as  in the case of charged helical stripes \cite{kanduc-epje},  the image dependent part of the interaction energy is non-zero and should be considered explicitly.

Note that physically the dielectric image-dependent part of the interaction energy corresponds to the interaction between the dielectrically induced image charge of the surface charge on the cylinder with the 
counterion as well as the image charge of the counterion with the surface charge on the cylinder, as follows from the general definition of the Green's function (\ref{eq:u_0u_im}). 

Using the above equations, we obtain the SC counterion number density profile ($\rho>a$) as
\begin{equation}
\tilde n(\rho) = C \rho^{-2\xi}\exp\Bigl[-\frac{\Xi}{\pi \xi} \,I(\rho/a)\Bigr], 
\label{nSC}
\end{equation}
where we have introduced the dimensionless integral 
\begin{equation}
I(x)=\sum_{m=0}^\infty\int_0^\infty\xi_m(t)K_m^2(tx)\,\rmd t,
\label{eqw}
\end{equation}
and the numerical prefactor $C$ is determined from the normalization condition, Eq.~(\ref{normrho}). 
We can extract two limiting behaviors for the function $I(x)$, viz.
\begin{equation}
I(x)\simeq\left\{
	\begin{array}{ll}
	\cfrac{\pi\Delta}{4(x-1)}&\quad x\to 1^+,\\
	&\\
	\cfrac{\pi^2\Delta(4+3\Delta)}{32(1+\Delta)x^3}&\quad x\to \infty.
	\end{array}
	\right.
\label{eqI}
\end{equation}
The limiting form of $I(x)$ for $x\to 1^+$ implies that at very small separations between a single counterion and the cylindrical surface, the dielectric self-energy has the same form as in the 
case of a counterion next to a planar wall, where the polarization effects can be described by an image charge inside the wall \cite{kanduc-epje}. 

The SC density for $\Delta=0$ reduces to the well-known form for a homogeneous system $n(\rho)\propto \rho^{-2\xi}$ \cite{Naji_CCT}.
If the dielectric core has a smaller dielectric constant than the medium ($\Delta >0$), as one typically encounters in the case of biological macromolecules, the image charges have the same sign as counterions and thus lead to a depletion of counterions in the vicinity of the cylinder. This behavior can be seen from the SC density profiles plotted in Fig.~\ref{fig:dens}a as the dielectric mismatch parameter $\Delta$ or the coupling parameter $\Xi$ is increased. They show a reduced density close to the cylinder (depletion zone) and a peak some distance away from the cylinder surface. 
One can estimate the location of the density peak by setting the density (\ref{nSC}) derivative to zero, $\rmd \tilde n(\rho)/\rmd \rho=0$. Using the approximation (\ref{eqI}) for $\rho=a+h$, where $h$ is the peak distance from the cylinder surface, we  obtain to the leading order for small  $h$ 
\begin{equation}
\Bigl(\frac{h}{a}\Bigr)^2\simeq\frac{\Delta\Xi}{8\xi^2},
\label{peak}
\end{equation}
which shows good agreement with the peak location values of the SC density profile as seen in  Fig.~\ref{fig:dens}a.

As noted before, effects of a finite counterion radius $R_c$ within the SC approximation can be taken into account exactly (due to the single particle nature of the theory)
by increasing the hard-core radius of the cylinder to  $a \to a+R_c$. The counterion-counterion excluded-volume repulsion effects are absent within the leading-order SC theory 
and enter only in the subleading terms. 

\subsection{Comparison with MC simulations}

We have performed extensive MC simulations in order to verify the validities of the weak- and strong-coupling approaches. The detailed description of MC simulation is given in Appendix \ref{app:MC}.

In Fig.~\ref{fig:dens}b we show the radial density profile of counterions around the cylinder without dielectric image effect, $\Delta=0$. Apparently in this case 
the simulation data  are always bracketed by the two analytical results obtained
from the WC and SC theories, Eqs.~(\ref{nPB}) and (\ref{nSC}), respectively. The WC theory is found to be valid 
at sufficiently small coupling parameter or sufficiently large radial distances from the cylinder. The SC theory is valid in the opposite regime, i.e., for sufficiently large coupling  parameters or sufficiently small radial distances. In fact, in agreement with the
general trend obtained for planar surfaces \cite{Netz,Naji,hoda}, the validity regime of the SC (WC) theory expands to large (smaller) separations
as the coupling becomes larger (smaller). In the case of a planar charged wall, the validity regime of the SC theory can be estimated systematically 
as $z/\mu < \sqrt{\Xi}$, where $z$ is the distance from the charged plane \cite{Netz}. In the case of a charged cylinder, a similar criterion is yet to be obtained. 

In Fig.~\ref{fig:dens}c, we show the results for the case where the dielectric constant of the cylinder is different from that of the medium such that 
$\Delta = 0.95$. As already discussed the dielectric discontinuities have no effects in the WC limit as confirmed also by the agreement obtained
between the simulation data and the WC (mean-field)  result for small $\Xi$. By contrast, the image effects become quite significant at large coupling parameters
as the density profile deviates qualitatively from those obtained with $\Delta=0$ (Fig.~\ref{fig:dens}b). 
The simulation data again show good agreement with the SC approximation at high enough couplings and small enough radial distances and in particular 
exhibit a depletion zone and a peak at small distances from the cylinder surface as predicted within the SC theory. Note that  in a dielectrically inhomogeneous system ($\Delta>0$) the SC theory has an explicit dependence on the coupling parameter $\Xi$  because of the self-interaction of counterions with their image charges.  

\section{Two like-charged cylinders}
\label{sec:two_cylinder}

\begin{figure}[t]
\centerline{(a)\psfig{figure=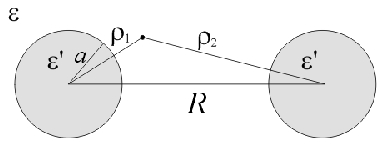,width=6cm}}
\centerline{(b)\psfig{figure=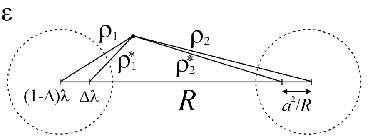,width=6cm}}
\caption{(a) Schematic top view of two identical charged dielectric cylinders oriented parallel in the $z$ direction. The system is confined
in a square box of lateral size $L_\bot$ containing the two cylinders and their neutralizing counterions (not shown).
(b) The potential generated by two charged dielectric cylinders can be reproduced equivalently by four linear (image) 
charges as explained in the text. 
}
\label{fig:model_2cyln}
\end{figure}

We now turn our attention to the problem of the interaction between
two like-charged cylinders, which is commonly used as a model 
for the interaction between rigid polyelectrolyte chains \cite{Oosawa}. We consider two identical parallel cylinders of radius $a$ oriented along the $z$ axis at an interaxial 
separation of $R$ (see Fig.~\ref{fig:model_2cyln}). The electric charge is assumed to be uniformly distributed on the surface for both cylinders with equal linear charge density $\lambda$. The cylinders are assumed to be infinitely long of length $L\to \infty$ and are standardly confined within a confining square box of lateral size $L_\bot$ (we shall assume
$L_\bot/a=60$ in Section \ref{subsec:wc_2cyln} and $L_\bot/a=100$ in Sections \ref{subsec:SC_2cyln}-\ref{subsec:SC_2cyln_im}), but we should emphasize
that in the regime of Manning parameters considered here, the lateral box size plays no significant role \cite{NajiNetzEPJE,Naji_CCT}. 

We follow the same approach as in the case of a single cylinder and thus first analyze the problem in the weak-coupling limit and then in the strong-coupling limit. We then 
compare the results from these two analytical theories with the MC simulations.

\subsection{Weak coupling limit: Mean-field theory}
\label{subsec:wc_2cyln}

The weak coupling regime ($\Xi\ll 1$) is again characterized by the mean-field PB equation. In this case however, no closed-form solution is available for the nonlinear PB theory.   We thus take full recourse to the numerical methods appropriate to the problem.  
In the two-cylinder geometry, the axial symmetry is broken but the translational symmetry along
the $z$ axis remains intact; hence one has to deal with a two-dimensional PB equation  in a finite bounding square $L_\bot\times L_\bot$, which is then solved numerically \cite{comsol}. 
We shall focus on the interaction force between the cylinders, which can be evaluated via the electrostatic stress tensor in a standard manner \cite{kanduc-helix}.
Note that the WC scheme used here based on the numerical solution of the PB equation allows for a full analysis of the dielectric discontinuity effects. (Later on in our SC analysis,
which will be based on analytical methods, we shall restrict our discussion to only the first-order image approximation, as a full analysis of
dielectric discontinuity effects in the SC limit is not yet available.) Here again the counterion size effect will be taken into account approximately by setting the distance of closest approach to each cylinder as  $a \to a+R_c$.  
Fluctuation contribution around the mean-field solution in cylindrical geometry is non-trivial to calculate \cite{Podgornik98} but is by construction small and we thus skip its detailed analysis. It does however depend crucially on the dielectric discontinuity at the cylinder surface.

\begin{figure}[t]
\centerline{\psfig{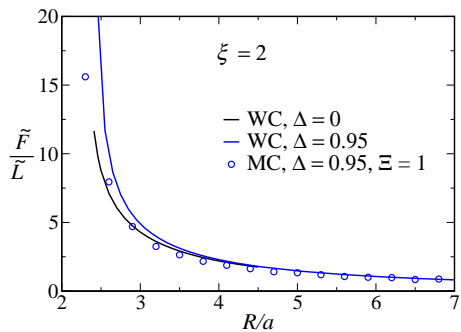}}
\caption{Rescaled force between two like-charged dielectric cylinders as a function of their interaxial separation. The WC prediction obtained by solving the
corresponding PB equation (solid lines for $\Delta=0$ and 0.95) are compared with the simulation data (symbols). }
\label{fig:forcedelta_wc}
\end{figure}

In general, the WC force obtained between two like-charged cylinders is repulsive and decays monotonically with their interaxial separation $R$. The mean-field prediction for the rescaled force per unit length defined via 
\begin{equation}
\frac{\widetilde F}{\widetilde L}\equiv \left(\frac{2\varepsilon\varepsilon_0}{\sigma_s^2\mu}\right)\,\frac FL,
\label{forceDef_WC}
\end{equation}
is shown in Fig.~\ref{fig:forcedelta_wc} as a function of the interaxial separation $R$ and 
for different values of the dielectric discontinuity parameter $\Delta$ (see Eq.~(\ref{eq:jump})). 
Note that in contrast to the case of a single cylinder studied in the previous Section, the dielectric discontinuity 
is expected to matter in the two-cylinder geometry even within the mean-field approximation. This is due to the axial symmetry breaking which leads to electric field penetrating into the cylindrical cores. As seen in the figure, the WC mean-field results  compare very well with the MC data (to be discussed later) at sufficiently large interaxial separations. The dielectric discontinuity has a relatively small effect on the interaction force. It leads to an increased repulsion (and also larger deviations from the mean-field prediction) 
at small interaxial separations but its effects diminish at larger  separations. 

\subsection{Strong coupling limit}
\label{subsec:SC_2cyln}

As noted before, the SC theory follows systematically from the leading order contribution to the partition function in the limit of infinite coupling parameter $\Xi\rightarrow \infty$ \cite{Netz}. 
The corresponding SC free energy involves only interactions between single 
counterions and the charged cylinders as well as the direct electrostatic interaction between the cylinders themselves (see Refs. \cite{Netz,NajiNetzEPJE}). In the 
presence of the dielectric discontinuity effects, the SC free energy per counterion can be written as 
\begin{equation}
\frac{\beta {\cal F}}{N}=\frac{\beta W_{00}}{N}-\ln
\int \exp\left[-\beta W_{\rm self}(\Av r)-\beta W_{0c}(\Av r)\right]\rmd\Av r.
\label{F1}
\end{equation}
Here $N$ is the total number of counterions and $W_{00}$ is the electrostatic energy due to the interaction between the two cylinders in the absence of any counterions, which 
contains both the direct Coulomb interaction between their surface charges as well as the contribution due to their images that results from the polarization of their 
dielectric cores.  Furthermore, $W_{0c}$ is the energy due to the interaction between a single counterion 
and the surface charges on both cylinders and includes contributions
from the corresponding image charges as well. Finally, $W_{\rm self}$ is the image self-energy of counterions, i.e., the contribution from the interaction of an individual 
counterion with its own image charges in both cylindrical cores. 
Note that the volume integration should be again taken  over the space available to counterions, i.e., inside the confining square box of lateral size $L_\bot$ excluding the
two cylindrical cores. Note also that on the SC level, the counterion excluded-volume repulsions are absent (as implied by the fact that strongly coupled counterions are highly isolated within large correlation holes in the SC limit) and only the excluded-volume interaction between the counterions and the cylinders will be present. In fact, the counterion size effects can be accounted for exactly within the SC theory via a hard-core repulsion which simply amounts to setting the effective (hard core) cylinder radius as $a+R_c$ \cite{kanduc-helix}.

As in the single cylinder case all charges are coupled with the interaction kernel, composed of direct and image part, but with the difference that now the image part of the kernel $u_{\rm im}^{(2)}(\Av r, \Av r')$ corresponds to two dielectric cylinders separated by the distance $R$. This kernel should satisfy the electrostatic boundary conditions on both cylindrical surfaces (\ref{bc}), which leads to very complicated numerical expressions and is unfortunately not available in a closed analytical form. Our aim is nevertheless to give a simple approximate analytical expression for the final results, so we need to make an approximation at this step. In what follows we focus on the first-order-image approximation, where we neglect higher orders of inter-cylindrical image interaction. This approximation implies that the image Green's functions of the two cylinders can be written as the sum of the Green's functions of the isolated cylinders, Eq.~(\ref{Vim}),
\begin{equation}
u_{\rm im}^{(2)}(\Av r, \Av r')\approx u_{\rm im}(\Av r_1, \Av r_1')+u_{\rm im}(\Av r_2, \Av r_2'),
\label{Vim2}
\end{equation}
where $u_{\rm im}(\Av r, \Av r')$ is the one-cylinder image kernel, Eq.~(\ref{Vim}). Here, $\Av r_1$, $\Av r_1'$ are distances centered at the first and $\Av r_2$, $\Av r_2'$ at the second cylinder,  for instance, 
\begin{equation}
\Av r_{1,2}=\Av r\mp(R/2)\Av e_x,
\end{equation} 
with $\Av e_x$ a unit vector pointing in the $x$ direction, i.e., horizontally on Fig.~\ref{fig:model_2cyln}. Recall that the surface charge is uniformly distributed on both cylinders so that the fixed 
charge distribution  is composed of two single cylinder distributions 
\begin{equation}
\sigma^{(2)}(\Av r)=\sigma(\Av r_1)+\sigma(\Av r_2),
\end{equation} 
where $\sigma(\Av r)$ is the charge distribution  of a single cylinder, Eq.~(\ref{n0}). With these assumptions one can then evaluate all interaction terms in Eq.~(\ref{F1}) analytically as follows. 

The cylinder-cylinder interaction energy per counterion can be obtained as 
\begin{eqnarray}
\frac{\beta W_{00}}{N}&=&\frac 12\,\frac{\beta}{N}\int\!\!\!\!\int \sigma^{(2)}(\Av r) u^{(2)}(\Av r,\Av r') \sigma^{(2)}(\Av r')\,\rmd \Av r\,\rmd \Av r',\nonumber\\
	&=&-\xi\,\ln\,R-\Delta \xi\,\ln\Bigl(1-\frac{a^2}{R^2}\Bigr).
\label{W00_delta}
\end{eqnarray}
Furthermore the image self-energy of a single counterion interacting with the dielectric cores of both cylinders can be derived in the form
\begin{eqnarray}
\beta W_{\rm self}(\Av r)&=&\frac 12 \beta(e_0q)^2 u_{\rm im}(\Av r_1,\Av r_1)+\frac 12 \beta(e_0q)^2 u_{\rm im}(\Av r_2,\Av r_2)\nonumber\\
	&=&-\frac{\Xi}{\pi \xi}\left[I(\rho_1/a)+I(\rho_2/a)\right],
\label{Wself}
\end{eqnarray} 
where $I(x)$ is defined in Eq.~(\ref{eqw}). Finally,  the single particle cylinder-counterion contribution can be written as the sum of the interaction with each cylinder separately and assumes the form 
\begin{eqnarray}
\beta W_{0c}(\Av r)&=&\beta e_0q \int u^{(2)}(\Av r,\Av r')\,\sigma^{(2)}(\Av r')\rmd \Av r',\nonumber\\
	&=&v(\Av r_1)+v(\Av r_2).
	\label{eq:W0c_two}
\end{eqnarray}
The cylinder-counterion energy contribution from the $\mu$-th cylinder (with $\mu=1,2$) is itself composed of three parts, 
\begin{equation}
v(\Av r_\mu)=v_0(\Av r_\mu)+v_{\rm same}(\Av r_\mu)+v_{\rm cross}(\Av r_\mu),
\label{eq:v_contrib}
\end{equation}
which correspond respectively to three parts of the Green's function, i.e., $u^{(2)}(\Av r, \Av r')\approx u_{0}(\Av r, \Av r')+u_{\rm im}(\Av r_1, \Av r_1')+u_{\rm im}(\Av r_2, \Av r_2')$, within the approximation implied by Eq.~(\ref{Vim2}) and will be evaluated as follows.

The first contribution in Eq.~(\ref{eq:v_contrib}),  $v_0(\Av r_\mu)$, is the direct interaction of the bare surface charge of each cylinder with a single counterion, i.e., for the the $\mu$-th cylinder, 
\begin{eqnarray}
v_0(\Av r_\mu)&=&\beta e_0 q\int u_0(\Av r_\mu,\Av r_\mu')\sigma(\Av r_\mu')\rmd \Av r',\nonumber\\
	&=&2\xi\,\ln\,\rho_\mu,
\end{eqnarray}
where $\rho_\mu$ is the radial distance from the center of the $\mu$-th cylinder to the counterion, and $\xi$ is the Manning parameter as defined in Eq.~(\ref{eq:xi_M}). 
The second contribution, 
\begin{equation}
v_{\rm same}(\Av r_\mu)=\beta e_0 q\int u_{\rm im}(\Av r_\mu,\Av r_\mu')\sigma(\Av r_\mu')\rmd \Av r'=0,
\end{equation}
corresponds to the interaction of a counterion image charge with the surface charge on the {\em same} cylinder, which is thus {\em zero} by symmetry reasons as  already discussed in the one-cylinder case, Eq.~(\ref{vself}). This contribution can be also considered as the interaction between the image of the surface charge on the same cylinder with the counterion.

Finally, we have the cross contributions due to 
the interaction of a counterion image in one cylinder with the surface charge on the {\em other} cylinder.
 Formally, it follows for each cylinder as (see Appendix \ref{app:cross}) 
\begin{eqnarray}
v_{\rm cross}(\Av r_\mu)&=&\beta e_0 q\int u_{\rm im}(\Av r_\mu,\Av r_\mu')\sigma(\Av r_\nu')\rmd \Av r',\nonumber\\
		&=&2\Delta \xi\,\ln\Bigl(\frac{\rho_{\mu}^*}{\rho_\mu}\Bigr),
\label{vcross}
\end{eqnarray}
for $\mu\ne \nu$,  where ${\rho_\mu^{*}}$ is the radial distance from an axis shifted by $a^2/R$ from the center of the $\mu$-th cylinder toward the axis of the other cylinder, see Fig.~\ref{fig:model_2cyln}b, i.e., 
\begin{equation}
{\rho_\mu^{*}}^2=\rho_\mu^2-2\,\Bigl(\frac{a^2}{R}\Bigr)\rho_\mu\cos\varphi_\mu+\Bigl(\frac{a^2}{R}\Bigr)^2, 
\end{equation}
or in cartesian coordinates, 
\begin{equation}
{\rho_{\mu}^*}^2=\Bigl(\frac{R}{2}-\frac{a^2}{R}\pm x\Bigr)^2+y^2. 
\end{equation}

The SC free energy thus follows by inserting the above expressions into Eq.~(\ref{F1}) in the form 
\begin{eqnarray}
\frac{\beta {\cal F}}{N}&=&-\xi\,\ln\,R-\Delta \xi\,\ln\Bigl(1-\frac{a^2}{R^2}\Bigr)-\nonumber \\
	&-&\ln\int\exp\biggl\{-\frac{\Xi}{\pi \xi}\Bigl[I\Bigl(\frac{\rho_1}{a}\Bigr)+I\Bigl(\frac{\rho_2}{a}\Bigr)\Bigr]- \nonumber \\
	&-&\xi(1-\Delta)\ln\,\rho_1^2\rho_2^2-
	\xi
	\Delta\,\ln\,{\rho_1^{*}}^2{\rho_2^{*}}^2\biggr\}\rmd\Av r.
\label{free}
\end{eqnarray}
The corresponding SC force between the two cylinders is obtained simply by differentiating the free energy with respect to $R$  
\begin{equation}
F=-\frac{\partial {\cal F}}{\partial R}, 
\label{free_to_force}
\end{equation}
which may be expressed  in dimensionless units 
according to Eq.~(\ref{forceDef_WC}). 

\begin{figure}[t]
\centerline{\psfig{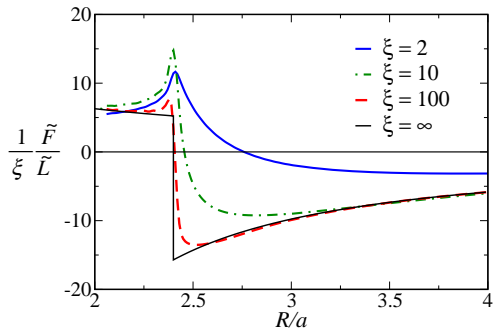}}
\caption{Rescaled SC force between two like-charged cylinders in the absence of a dielectric discontinuity ($\Delta=0$) as a function of the rescaled interaxial separation for various Manning parameters as indicated on the graph. }
\label{fig:force0}
\end{figure}

The SC density profile follows from the general expression (\ref{eq:density_ci}), which is the same as the integrand in (\ref{free}), viz. 
\begin{eqnarray}
\tilde n(\Av r) = C \exp\biggl\{&-&\frac{\Xi}{\pi \xi}\Bigl[I\Bigl(\frac{\rho_1}{a}\Bigr)+I\Bigl(\frac{\rho_2}{a}\Bigr)\Bigr]- \label{dens2}\\
	&-&\,\xi(1-\Delta)\ln\,\rho_1^2\rho_2^2-
	\xi
	\Delta\,\ln\,{\rho_1^{*}}^2{\rho_2^{*}}^2\biggr\}\nonumber.
\end{eqnarray}
where the normalization prefactor is determined from the electroneutrality condition.

Note that the forms of the SC free energy, Eq.~(\ref{free}), and the SC counterion density profile, Eq.~(\ref{dens2}), reflect  the fact that, within the first-order image approximation (and using the appropriate volume constraints for counterions), the two charged dielectric cylindrical cores can be equivalently replaced by four parallel lines of charges, two of which 
are located along the central axis for each of the cylinders with (renormalized) linear charge density   $(1-\Delta) \lambda$ and the two other lines of charges with linear charge density $+\Delta \lambda$
are placed along two axes shifted by $a^2/R$ from the central axis for each of the cylinders, see Fig.~\ref{fig:model_2cyln}. 
Note that the image charge language is applicable for two charged dielectric cylinders in the absence of pointlike charges (counterions) \cite{smythe}, and can be exactly solved by applying higher orders of images. But in order to be consistent with the first-order image counterion-cylinder approximation, we have used the same image kernel (\ref{Vim2}), which accounts for the first-order image approximation between the two cylinders as well (\ref{F1}).

\begin{figure}[t]\begin{center}
	\begin{minipage}[b]{0.27\textwidth}\begin{center}
		\includegraphics[width=\textwidth]{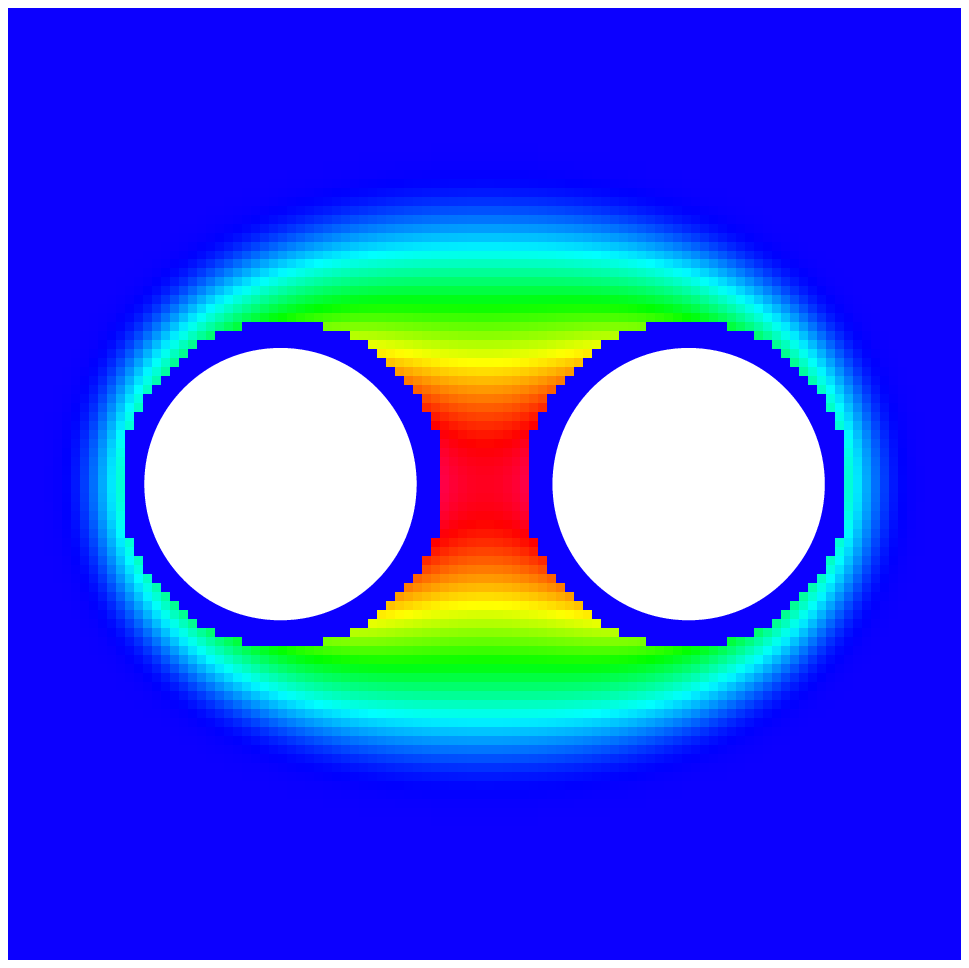}
	\end{center}\end{minipage} \hskip0.25cm
	\begin{minipage}[b]{0.082\textwidth}\begin{center}
		\includegraphics[width=\textwidth]{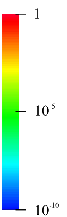}
	\end{center}\end{minipage} \hskip0.25cm
\caption{Counterion density around two parallel like-charged cylinders with Manning parameter $\xi=10$
		at interaxial separtaion $R=3a$ and in the absence of image charges ($\Delta=0$). 
		The density is shown across an arbitrary plane perpendicular to the  two cylinders in a color-coded fashion as indicated on the graph. 
                The small depletion zone around each cylinder is due to the counterion volume exclusion as counterions are assumed to have a finite radius of $R_c=0.2a$. }
		\label{fig:densdelta0}
\end{center}\end{figure}

\begin{figure*}[t]
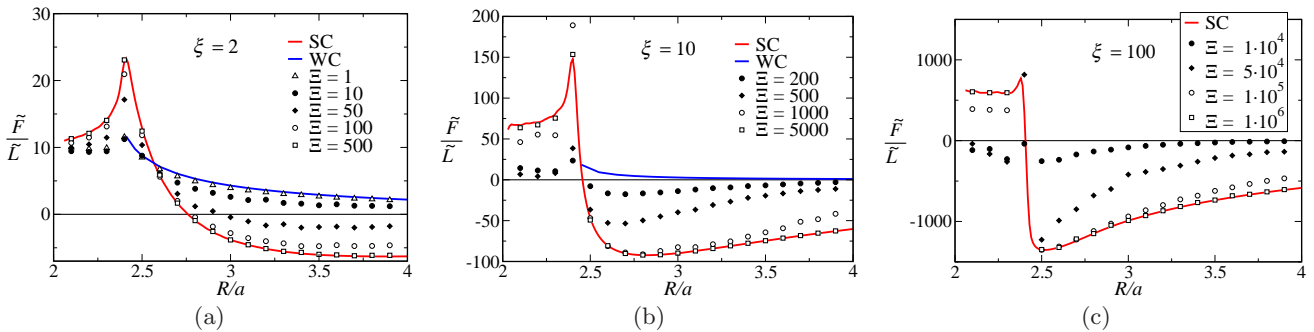
\begin{center}
	\begin{minipage}[b]{0.302\textwidth}\begin{center}
		\includegraphics[width=\textwidth]{forceQ2.eps} (a)
	\end{center}\end{minipage} \hskip0.25cm
	\begin{minipage}[b]{0.31\textwidth}\begin{center}
		\includegraphics[width=\textwidth]{forceQ10.eps} (b)
	\end{center}\end{minipage} \hskip0.25cm
	\begin{minipage}[b]{0.313\textwidth}\begin{center}
		\includegraphics[width=\textwidth]{forceQ100.eps} (c)
	\end{center}\end{minipage} \hskip0.25cm
	\caption{Rescaled interaction force between two like-charged cylinders in the absence of image charges
	 ($\Delta=0$)  as a function of the rescaled interaxial separation for  Manning parameters $\xi=2$ (a), 10 (b),  100 (c) and
	different values of the coupling parameters as indicated on the graphs. The symbols represent the simulation data, the red lines the prediction of the 
	SC theory (Eq.~(\ref{free0})) and the blue lines are that of the WC theory (Section \ref{subsec:wc_2cyln}). 
	}
	\label{fig:force0MC}
\end{center}\end{figure*}

\subsection{SC interaction in a dielectrically homogeneous  system  ($\Delta=0$)}

In order to proceed, let us first consider the case where there is no dielectric mismatch between the cylinders and the surrounding medium, i.e., $\Delta=0$; the image charges will thus  be absent. The SC interaction in this case has been considered also in previous works  \cite{NajiNetzEPJE, NajiArnold, arnoldholm}. We shall reproduce some of these previous results for the sake of completeness, but will also provide several new results, including a direct comparison with MC simulations for the force dependence as a function of separation at different values of the coupling parameter as well as a global interaction phase diagram, which have not been considered previously. 

In the absence of image charges, the SC free energy (\ref{free}) reduces to a simple form as
\begin{equation}
\frac{\beta {\cal F}}{N}=-\xi\,\ln\,R-\ln\int\exp\left(-2\xi\,\ln\,\rho_1\rho_2\right)\rmd\Av r, 
\label{free0}
\end{equation}
where the first term is again the bare repulsion between the two cylinders and the second term contains energetic and entropic contributions from counterions. The results for the  SC force that follow from Eq.~(\ref{free0}) are shown in Fig.~\ref{fig:force0} for a few different Manning parameters, where we have accounted for the finite 
counterion radius by choosing $R_c=0.2 a$. The SC force is repulsive at small separations and becomes attractive beyond the equilibrium interaxial distance $R^*$ where the force vanishes. The SC attraction is mediated by counterions that are strongly coupled to both cylinders in this limit and are thus accumulated mainly in the region between the two cylinders as can be seen  
from the counterion distribution (\ref{dens2}) in Fig.~\ref{fig:densdelta0}. 
This is a direct consequence of the energy  contributions included in the second term in Eq.~(\ref{free0}) \cite{NajiNetzEPJE, NajiArnold}. 

By decreasing the separation between the cylinders, the entropic osmotic pressure from counterions sandwiched between the two cylinders 
becomes increasingly important and the effective interaction becomes repulsive.  This repulsion is reduced as the surface-surface separation, $R-2a$, becomes smaller than the counterion diameter, $2R_c$ as the counterions are depleted from the intervening region due to excluded-volume effects. This behavior due to the presence of the counterion depletion interaction has been investigated thoroughly in the case of cylinders with helical charge pattern and we shall not delve on it any further here \cite{kanduc-helix}.

As seen in Fig.~\ref{fig:force0} the entropic repulsion effects at small  separations decrease when the Manning parameter $\xi$ is increased. 
For very large Manning parameters, the free energy is dominated by purely energetic contributions and does not contain any temperature effects. In fact the limit $\xi\rightarrow \infty$ formally corresponds to the zero temperature limit as we have already taken the limit $\Xi\rightarrow \infty$ appropriate within the SC approximation. The connection between the SC limit and the zero temperature limit has been analyzed in detail in Refs. \cite{hoda, arnoldholm}.
The asymptotic form of the SC force in the limit of large $\xi$ can be obtained as 
\begin{equation}
\widetilde F/\widetilde L=\left\{
	\begin{array}{cl}
	\cfrac{4\pi \xi}{R/a}\,&	\quad R<2(a+R_c),\\
	&\\
	-\cfrac{4\pi \xi}{R/a}\Bigl(\cfrac{2R}{R-a-R_c}-1\Bigr)&	\quad  R>2(a+R_c),
	\end{array}
	\right.
\end{equation}
which is shown as a black solid line in Fig.~\ref{fig:force0}. The equilibrium hard-core surface-surface separation
\begin{equation}
\delta R^*\equiv R^*-2(a+R_c), 
\end{equation}
tends to zero as $\delta R^*/a \simeq 2/(3\xi)$ as $\xi\rightarrow \infty$ \cite{NajiNetzEPJE}. 
Note that even for moderate values of the  Manning parameters, $\xi\sim 5$, the SC theory 
predicts a closely packed {\em bound state} with a relatively small surface-surface separation $\delta R^*/a \sim 0.1$. 

Let us now consider the situation where the coupling parameter has a finite value. In this case we shall study the system using MC simulations as detailed in Appendices \ref{app:MC} and \ref{app:MC_force}. The MC results for the interaction force between like-charged cylinders (with no dielectric mismatch) are shown in Fig.~\ref{fig:force0MC} along with the WC and SC predictions. As seen in Fig.~\ref{fig:force0MC}a, the WC results agree with the simulation data for a sufficiently small coupling parameters $\Xi\lesssim 1$ as expected. Furthermore, by increasing the coupling parameter, the MC results start deviating strongly from the WC prediction and tend to the SC prediction,
where a reasonable agreement is obtained already at coupling parameters of the order $\Xi\sim 100$.  Note that the WC and SC predictions bracket the simulation data and thus establish upper and lower bounds for the interaction force at any realistic value of the coupling parameter.

\begin{figure*}[t]\begin{center}
\begin{minipage}[b]{0.21\textwidth}\begin{center}
\includegraphics[width=\textwidth]{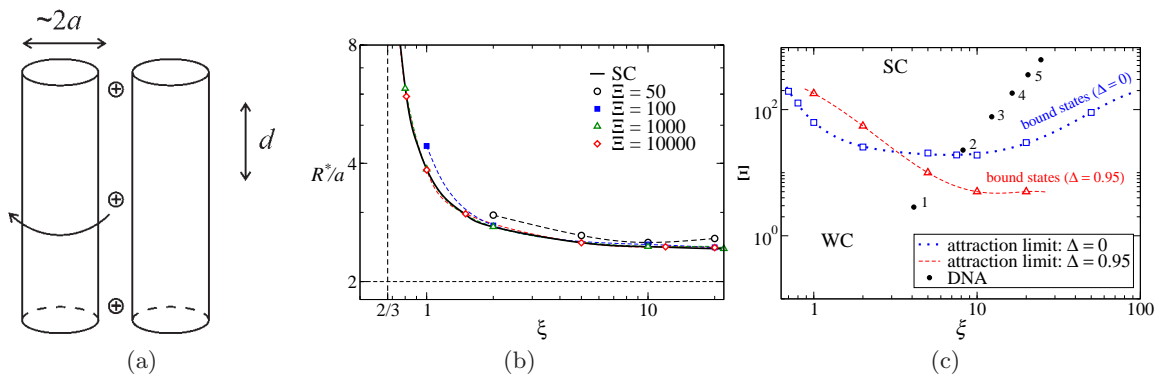} (a)
\end{center}\end{minipage} \hskip0.25cm
\begin{minipage}[b]{0.31\textwidth}\begin{center}
\includegraphics[width=\textwidth]{boundQa.eps} (b)
\end{center}\end{minipage}
\begin{minipage}[b]{0.31\textwidth}\begin{center}
\includegraphics[width=\textwidth]{diagramXiQ2.eps} (c)
\end{center}\end{minipage}
\caption{(a) Schematic representation of  the validity  criterion for the SC theory at large $\xi$. 
The SC predicts the configuration where all counterions are located between the cylinders. This argument fails if counterion-counterion repulsion is too strong.
(b) Bound-state interaxial separation as a function of the Manning parameter. The solid line shows the SC prediction and the symbols represent the simulation data. 
The horizontal dashed line represents the closest approach distance between the two cylinders, $2a$, and the 
vertical dashed lines represents the minimum value of the Manning parameter  $\xi^*=2/3$ where SC force may be attractive. 
(c) Phase diagram indicating the repulsion and attraction between two like-cylinders with no image charges in terms of $\Xi$ and $\xi$
in the absence (blue squares) and in the presence of image charges (red triangles with $\Delta=0.95$).   
Symbols show the boundary line obtained from MC simulations.  Filled circles show the points appropriate for DNA with counterions of valencies in the range $q=1, \ldots, 6$. 
In (b) and (c) dashed lines are guides to the eye. }
\label{fig:diagrams0}
\end{center}\end{figure*}

In Figs.~\ref{fig:force0MC}b and c we show the results for larger Manning parameters of $\xi=10$ and 100, respectively. As Manning parameter becomes larger, a larger $\Xi$ is required in order to achieve the same level of agreement with the SC prediction at a given separation and thus  the convergence to the SC limit becomes weaker. 
This effect is more pronounced at larger interaxial separations. In fact, as is generally known \cite{Netz}, 
the SC effects become more dominant and the SC theory becomes increasingly more accurate 
at smaller surface-surface separations.  In general, the applicability regime
of the SC theory (obtained in the limit of $\Xi\rightarrow \infty$) to the situations where the coupling parameter $\Xi$ is finite  can be specified 
via simple validity criteria as applied successfully in several previous studies \cite{Netz, asim,Naji,NajiNetzEPJE,arnoldholm,jho-prl,hoda}. In what follows we shall derive the
particular form of the SC validity criterion for the case of two like-charged cylinders.

It should be noted first that the SC theory (obtained as an exact asymptotic theory in the limit of $\Xi\rightarrow \infty$) contains only contributions stemming from the interaction of individual counterions with charged objects and thus the counterion-counterion repulsions and other higher order many-body effects are absent. These effects however become increasingly more important as the coupling parameter is decreased, depending crucially on the precise value of all the system parameters. In the case of two cylinders, as noted before, counterions are mostly accumulated in the intervening region between the two cylinders  (Fig.~\ref{fig:densdelta0}) where the typical counterion spacing can be estimated by stipulating the local electroneutrality condition as $d = qe_0/(2\lambda)$ or in rescaled units as $d/\mu = \Xi/2\xi$. It is thus evident that as the Manning parameter  becomes larger at a fixed coupling parameter the spacing between counterions tends to become small. Hence, a larger effect due to counterion-counterion repulsions and thus larger deviations from the SC theory would be expected to be observed at larger $\xi$. 
In other words, the SC theory is expected to overestimate the counterion density in the intervening region at any {\em finite} value of the coupling parameter, which is consistent
with the general result that the SC theory gives the upper bound for the density profile of the counterions \cite{asim,hoda,Netz,Naji}.  
One can thus identify the validity regime of the SC theory by estimating the counterion-counterion contributions and comparing them with the sinlge-particle counterion-cylinder contribution at a given interaxial separation. This can be done by noting that the ground state configuration (obtained for $\xi \rightarrow \infty$ and $\Xi\rightarrow \infty$) 
is predicted within the SC theory to be the configuration where all counterions are localized between
the two cylinders and are lined-up along the $z$ axis as obviously  favored by the electrostatic interaction energy.

When the coupling parameter is reduced to a finite value, repulsions between counterions  become more important and some counterions tend to escape from the intervening region to the external side of the cylinders (shown by an arrow in Fig.~\ref{fig:diagrams0}a). In order to prevent this from happening, the coupling parameter $\Xi$ has to be large enough. By comparing the energy of the ground state with the energy of a single defect obtained by allowing one counterion to move to the external side of one of the cylinders, it follows analytically that for a large $\xi$ and in the absence of image charges, one should have 
\begin{equation}
	\begin{array}{ll}
	\Xi>4.2\,\xi^2	&\quad R\sim 2a,\\
	&\\
	\Xi>2.9\,\xi^2 \Bigl(\cfrac Ra\Bigr)& \quad R\gg 2a.\\
	\end{array}
\label{SCcriterion}
\end{equation}
These criteria indeed capture the trends observed in Fig.~\ref{fig:force0MC} 
when viewed in terms of the three main parameters describing the system, i.e., $\Xi$, $\xi$ and $R$. 
They  automatically cover the one proposed before based on the surface-surface separation between the cylinders, where one requires that the surface-surface separation between the cylinders must be smaller than the spacing between counterions, $\delta R<d$, or in rescaled units  \cite{NajiNetzEPJE, NajiArnold}
\begin{equation}
\delta R/\mu<\Xi/(2\xi).
\label{eq:crit_old}
\end{equation}
This latter criterion is inspired by the observations in the case of charged planar walls where it can indeed  be derived from systematic analysis of higher-order 
corrections to the SC theory  \cite{Netz}. It can describe the validity regime of the SC predictions for the equilibrium separation between strongly coupled cylinders that form
a closely packed bound state \cite{NajiNetzEPJE, NajiArnold}, but when the applicability of the SC theory for the interaction force is considered, the criteria (\ref{SCcriterion}) should be used instead. The difference between the new criteria (\ref{SCcriterion})  and the one  in Eq.~(\ref{eq:crit_old}) is in fact related to  the fundamental difference  between the topology of the space available to counterions in the case of two cylinders as compared to  two planar walls: in the latter case counterions can not ``escape" from the intersurface gap 
and the ground state of the system has a universal two-dimensional configuration \cite{Naji,hoda,Netz}, while in the former the ground state corresponds to a 
one dimensional arrangement of counterions and the whole space is available for thermal excitations from the ground state.

The criteria  (\ref{SCcriterion}) are more stringent and cover also the situation where the
cylinders are placed at large interaxial separations.  The ground state of neutralizing counterions between two charged cylinders (without the effects of the dielectric mismatch)  has been 
investigated extensively  by Arnold and Holm in Ref. \cite{arnoldholm}, where  a validity criterion for 
the SC theory has been proposed   as $\Xi>3.45\,\xi^2 (R/a)$ (when expressed in rescaled units) based on computer simulations. This 
agrees with our estimate in Eq.~(\ref{SCcriterion}) within 16\% of the numerical prefactor.

Let us now turn our attention to the behavior of the equilibrium interaxial separation or the so-called bound-state separation, $R^*$, where the interaction force between the two cylinders vanishes. As noted before, the SC theory is generally more accurate at smaller separations. A closer inspection of Fig.~\ref{fig:force0MC} shows that the SC prediction for the bound-state separation agrees with the MC data even outside the regime set by the SC criteria (\ref{SCcriterion}), which reiterates the point discussed above that a less stringent  criterion such as Eq.~(\ref{eq:crit_old}) would be sufficient to describe the validity of the SC prediction for the bound-state separation. 
In  Fig.~\ref{fig:diagrams0}b the bound-state interaxial separation is plotted as a function of the Manning parameter, which shows that a reasonable agreement between MC data and the SC theory for this quantity can be achieved for a coupling parameter as small as $\Xi\simeq 50$, in agreement with previous results in Ref. \cite{NajiArnold}. 
As seen in the figure, the bound-state separation diverges (i.e., the cylinders unbind) for Manning parameter approaching a minimum value of $\xi^*=2/3$ \cite{NajiNetzEPJE} below
which the two cylinders merely repel each other. 

Figure \ref{fig:diagrams0}b can be also thought of as a phase diagram identifying the attraction (the region above the lines) and repulsion (the region below the lines) regimes between two like-charged cylinders in terms of the parameters $R^*$ and $\xi$ for a given coupling parameter $\Xi$. 

In Fig.~\ref{fig:diagrams0}c we show a global phase diagram in terms of the parameters $\Xi$ and $\xi$. The boundary lines are determined from MC simulations (open symbols) and separate the region  (above the lines) where the interaction force becomes  attractive at some finite interaxial separation between the cylinders and the region  (under the lines) where the force never becomes attractive at {\em any} interaxial separation. In other words, the boundary lines themselves correspond to the parameter values where the force-distance diagrams exhibit 
a global minimum at $F=0$, i.e. the force curve only touches the abscissa. 
They were determined by bisection procedure for a series of MC runs. If the cylinder parameter values are chosen to describe DNA, we have $\xi=4.1\,q$ and $\Xi=2.8\,q^3$ for $q$ valency counterions, which are shown as solid circles in Fig.~\ref{fig:diagrams0}c. This suggests that DNA-DNA interaction in the case of monovalent counterions falls well within the repulsive region, while with divalent counterions and beyond one can expect to observe an attractive force, in qualitative agreement with recent experiments \cite{rau-1,rau-2, besteman}. 

\begin{figure}[t]
	\begin{center}
		\begin{minipage}[b]{0.2\textwidth}\begin{center}
			\includegraphics[width=\textwidth]{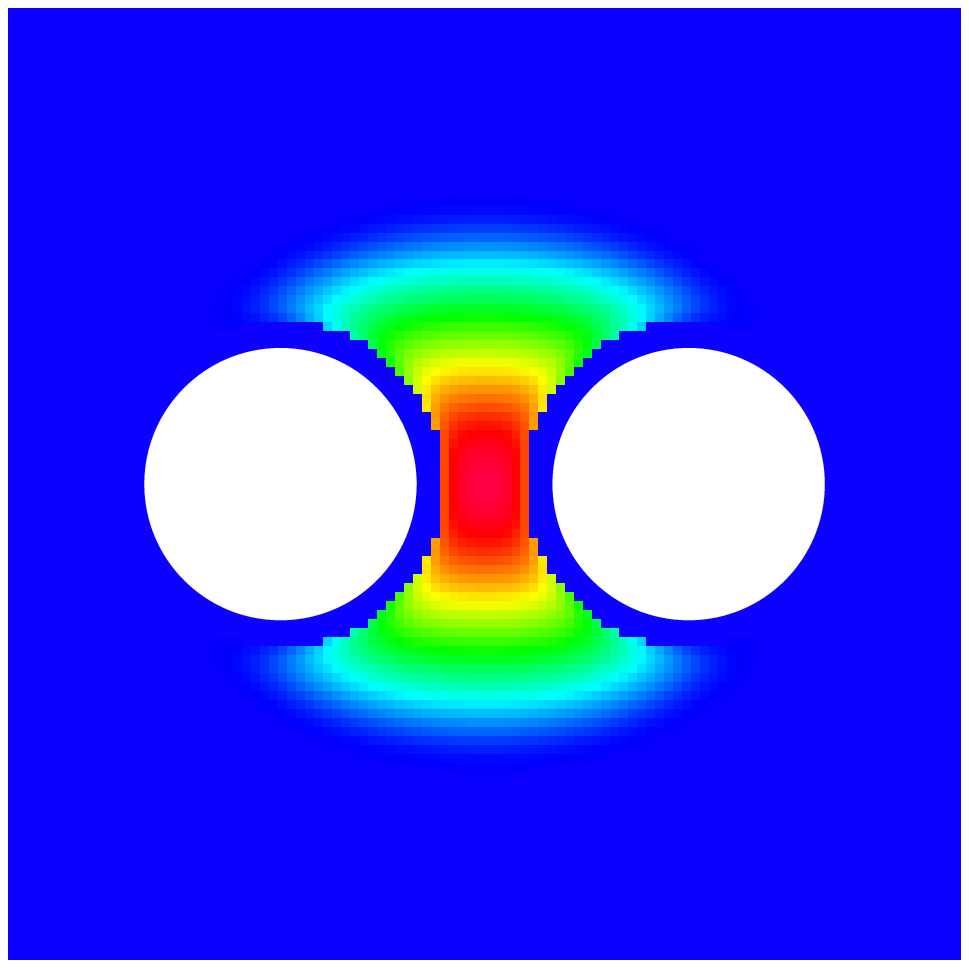} (a) $\Xi=100$
			\includegraphics[width=\textwidth]{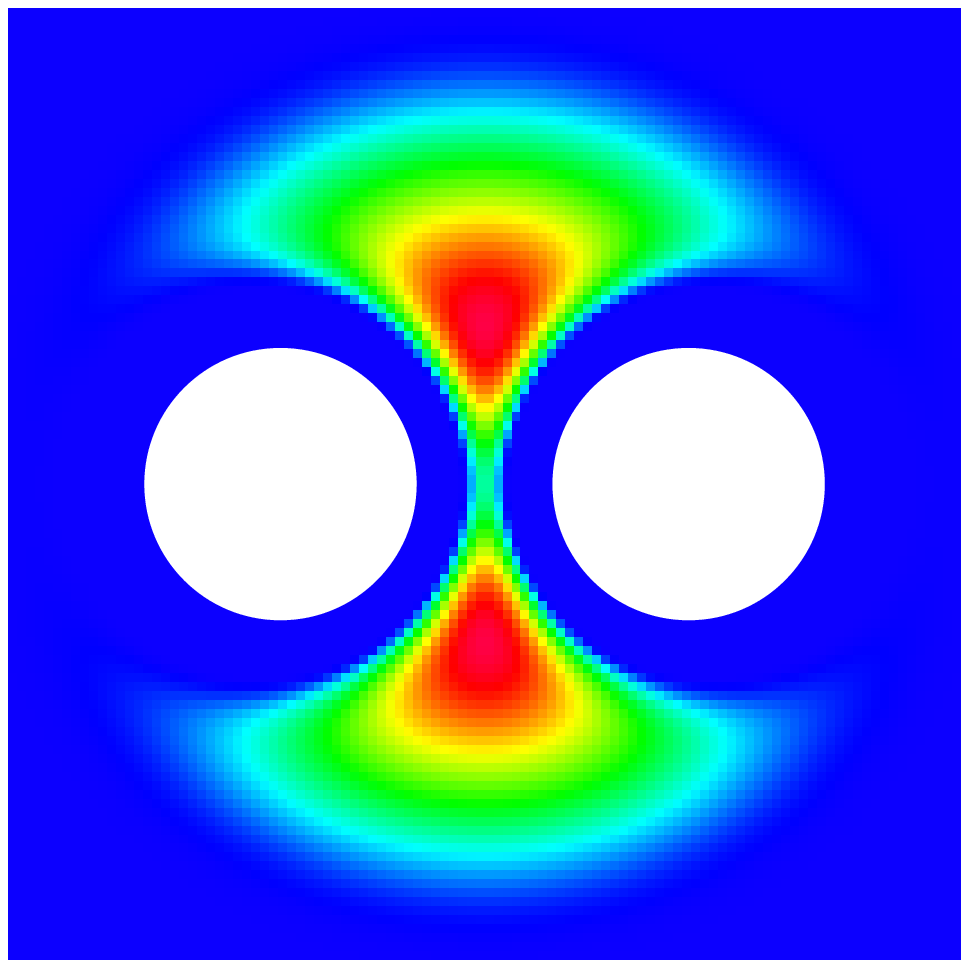} (c) $\Xi=1000$
		\end{center}\end{minipage} \hskip0.25cm
		\begin{minipage}[b]{0.2\textwidth}\begin{center}
			\includegraphics[width=\textwidth]{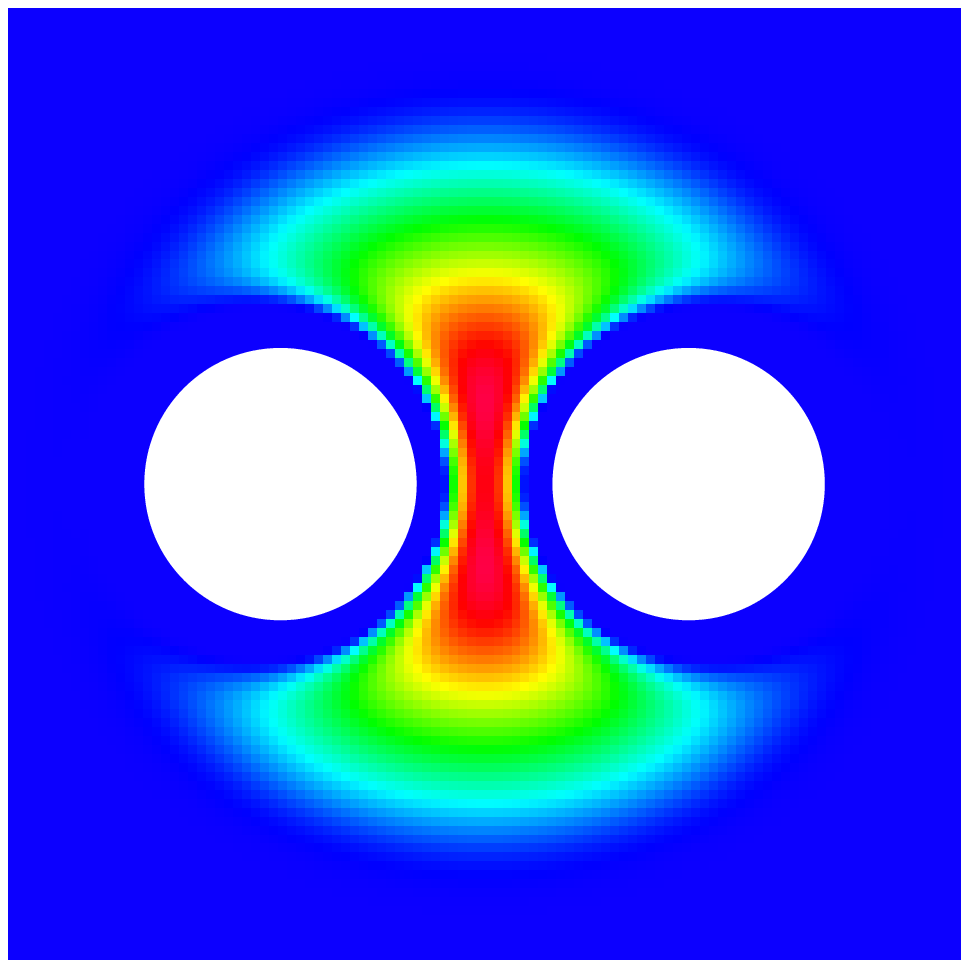} (b) $\Xi=500$
			\includegraphics[width=\textwidth]{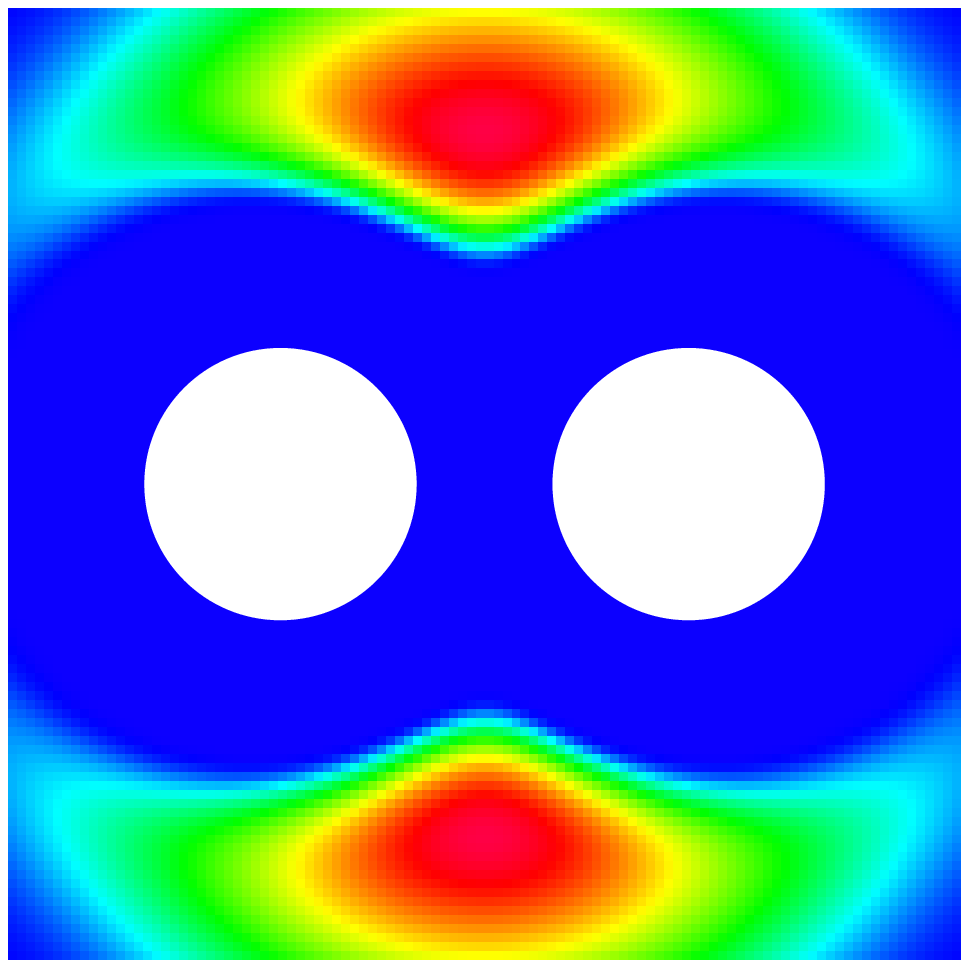} (d) $\Xi=5000$
		\end{center}\end{minipage}
		\caption{
		Counterion density around two parallel like-charged cylinders with Manning parameter $\xi=10$ and the dielectric discontinuity parameter $\Delta=0.95$
		at interaxial separtaion $R=3a$. The density is shown across an arbitrary plane perpendicular to the  two cylinders in a color-coded fashion as indicated on the graph. 
                The small depletion zone around each cylinder is due to the counterion volume exclusion as counterions are assumed to have a finite radius of $R_c=0.2a$. A large
                depletion zone develops due to the influence of image charges as the coupling parameter increases from $\Xi=100$ to 5000 (a-d). 
}	
		\label{fig:densdelta}
	\end{center}
\end{figure}

\begin{figure*}[t]
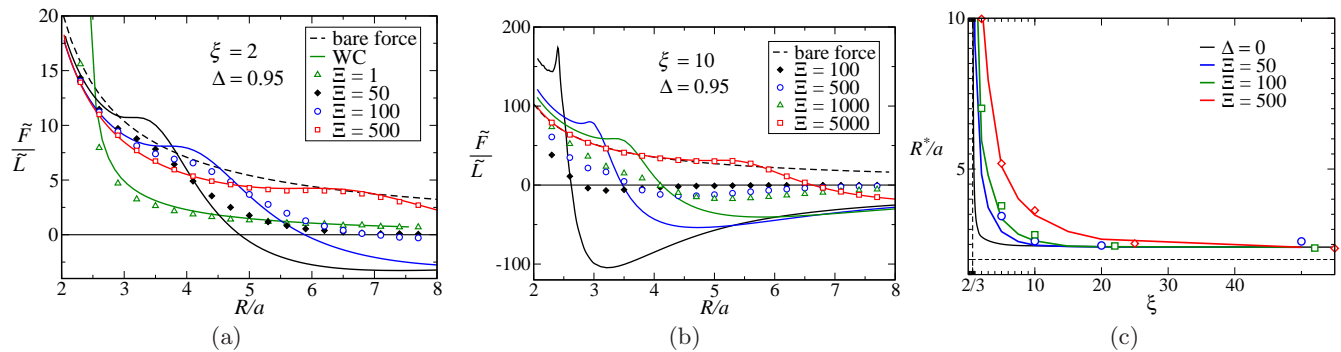
\begin{center}
\begin{minipage}[b]{0.32\textwidth}\begin{center}
\includegraphics[width=\textwidth]{forceQ2Delta.eps} (a)
\end{center}\end{minipage} \hskip0.25cm
\begin{minipage}[b]{0.32\textwidth}\begin{center}
\includegraphics[width=\textwidth]{forceQ10Delta.eps} (b)
\end{center}\end{minipage}
\begin{minipage}[b]{0.32\textwidth}\begin{center}
\includegraphics[width=\textwidth]{boundQdelta.eps} (c)
\end{center}\end{minipage}
\caption{
(a) and (b) Force results between two dielectric charged cylinders ($\Delta=0.95$) for various coupling parameters $\Xi$. 
The MC results (symbols) are compared with the corresponding (same color) SC results obtained from Eq.~(\ref{free}) and shown here as solid lines. In (a) the
MC results for $\Xi=1$ are compared with WC prediction (solid green line) as discussed in Section \ref{subsec:wc_2cyln}. The dashed lines shows the bare
electrostatic interaction between the two cylinders in the absence of counterions,  Eq.~(\ref{F00}) in the appendix. 
(c) Bound-state distance for two cylinders with $\Delta=0.95$. Here MC results (symbols) are compared with the SC predictions (solid lines of the same color).}
\label{fig:forceQdelta}
\end{center}\end{figure*}

\subsection{Image-charge effects  ($\Delta>0$) in the SC limit}
\label{subsec:SC_2cyln_im}

We now consider the case of two charged cylinders with a dielectric core whose dielectric constant may be in general different from that of the surrounding medium. In this case, the induced image charges (due to the polarization of the dielectric cores) can  play an important role as well. However, as we discussed before (Fig.~\ref{fig:forcedelta_wc}), the image-charge effects turn out to be small in the WC regime where the coupling parameter is small (e.g., when counterions are  monovalent). This may be the reason why such dielectric effects have not yet been throughly investigated in the particular case of two charged cylinders. The situation turns out to be quite different in the opposite limit of strong coupling as we will see in this Section. This is tied in with  a fundamental 
difference between counterion distribution in the SC limit as opposed to the WC limit: while in the former limit one deals with highly isolated single counterion close to
a charged surface, in the latter limit individual nature of counterions fades in the wake of dominant collective many-body effects (as counterions form a diffuse and uncorrelated ionic cloud 
around the cylinders) and thus the polarization effects are highly reduced. 

If the dielectric cores have a smaller dielectric constant than the medium ($\Delta >0$), as relevant to most macromolecules in water, the induced image charges will 
have the same sign as the counterions and thus tend to cause depletion of counterions from the vicinity of the dielectric cores. This 
behavior is shown in Fig.~\ref{fig:densdelta}, where the SC density is plotted across an arbitrary plane perpendicular to the  two cylinders in a color-coded fashion. 
As can be seen, the depletion effect is enhanced as the coupling parameter, $\Xi$,
is increased such that a large  ``depletion zone" develops that eventually encloses both cylinders. (Note that 
the SC free energy in the presence of a dielectric discontinuity, Eq.~(\ref{free}), has an explicit dependence on the coupling parameter $\Xi$ as already known from the case 
of planar charged dielectric slabs  \cite{kanduc-epje}). 

In fact, the image depletion at larger couplings turns out to have a similar effect qualitatively as the counterion volume exclusion, i.e., as if the counterion radius $R_c$ is renormalized
to a large effective value. Note that the region with highest concentration (shown in red color) is  ``squeezed" and eventually splits in two regions shifted away from the common $x-z$ plane that passes through the axes of the two cylinders. The splitting appears when the second derivative of the density (\ref{dens2}) with respect to the $y$ coordinate becomes positive at the midpoint, $\partial^2 \tilde n(\Av r)/\partial y^2\vert_{y=0}>0$. Using the small-distance approximation $\rho\to a^+$ from Eq.~(\ref{eqI}), we can estimate the distance $R$ at which the density splitting happens as 
\begin{equation}
\Bigl(\frac{R-2a}{a}\Bigr)^2\sim\frac{\Delta \Xi}{2(1+3\Delta)\xi^2}. 
\end{equation}
Thus for the Manning parameter $\xi=10$ and at separation $R=3a$, the splitting is estimated to occur at $\Xi\sim 800$ as is the case in Fig.~\ref{fig:densdelta}.
	
In Figs.~\ref{fig:forceQdelta}a and b, we show the SC force per unit length (solid lines) as obtained from Eqs.~(\ref{free}) and (\ref{free_to_force}) along with 
the MC simulation results (symbols) for two different Manning parameters and several different values of the coupling parameter. 
As seen upon increasing the coupling parameter, the repulsive peak at small separations shifts towards larger values of the interaxial separation $R$, which reflects 
the emergence of a depletion zone  due to  repulsions between counterions and their image charges as discussed before.  As a result, the  main contribution to the inter-cylinder 
interaction  at small separations comes from the bare interaction between the two cylinders and the force is approximately given by the bare repulsion force $F_{00}$ as given 
by Eq.~(\ref{F00}) in the appendix (dashed line, Fig.~\ref{fig:forceQdelta}b). 
At larger separations, we still find strong SC-type attraction even in the presence of image charges. 
One should thus note that in general the repulsive forces are enhanced in the effective interaction in the SC limit especially at small separations 
as compared to a homogeneous system with no image charges.
Therefore,  unlike in the WC limit (see Fig.~\ref{fig:forcedelta_wc}), the introduction of image charges may lead to qualitative changes in the behavior for a strongly coupled system. 

In general,  the SC results for the interaction force show better agreement with the simulations for larger $\Xi$ and smaller separation distances $R$. 
Especially the  results for the bound-state interaxial separation show very good agreement between  the SC results (solid lines) 
and the MC data (symbols)  as shown in Fig.~\ref{fig:forceQdelta}c (compare with Fig.~\ref{fig:diagrams0}b). 

In plan-parallel geometry the repulsive self-image interaction compresses counterions toward the mid-plane and hence expands the range of the SC regime \cite{jho-prl}. Here in the cylindrical geometry such mechanism cannot work, since counterions can escape from inter-cylindrical region. Therefore, it seems that dielectric image effects do not have a noticable influence on the range of validity of the SC approximation.

Note that just as in the case with $\Delta=0$ (blue line in Fig.~\ref{fig:diagrams0}c), the simulation results show no attraction between the cylinders at too low electrostatic coupling parameters. 
The attraction can only appear once the coupling parameter exceeds a threshold value. This threshold value is shown as a red line in the global phase diagram
in Fig.~\ref{fig:diagrams0}c for the dielectric discontinuity parameter $\Delta =0.95$. 
Note that bare cylinder-cylinder repulsion is enhanced in the case of images ($\Delta>0$), eq.(\ref{F1}).
Therefore, larger coupling parameter $\Xi$ is needed
to get attraction at smaller Manning parameters compared to no-image case (red line stays above blue for $\xi<3$ and $\Delta=0.95$ in Fig.~\ref{fig:diagrams0}c). 
For larger Manning parameters, the attractive interaction between the counterion image and the cylindrical charges (represented by the term $v_\textrm{cross}$, Eq.~(\ref{vcross})) becomes 
more important and, hence, attraction can appear even at lower coupling parameters
(red line is below the blue line for $\xi>3$ on Fig.~\ref{fig:diagrams0}c).

\section{Conclusions}

We have analyzed the electrostatic interaction between two like-charged cylindrical macromolecules surrounded by counterions. We have in particular examined 
the role of image charges when the cylinders have a dielectric constant which is different from that of the surrounding medium within the 
the weak- and strong-coupling frameworks elaborated first by Netz and coworkers \cite{Naji,hoda,Netz}. 
The two limits are defined in terms of a single coupling parameter $\Xi$. 

The weak coupling limit, or equivalently the mean-field theory, is relevant when the valency of counterions
and/or the macromolecule surface charge density is small and/or when the dielectric constant of the medium is large enough. It is based on the Poisson-Boltzmann equation which turns out
to give purely repulsive interactions between two-like charged cylinders. The image-charge effects in this case turn out to be small even at small separations and quickly diminish
as the inter-cylinder separation is increased.

On the contrary, the strong coupling limit is relevant when the valency of counterions is large and/or the macromolecule surface charge density  is large
and/or when the dielectric constant of the medium is small enough. The SC approach is based on a single-particle description which is obtained as an exact limiting result
for large coupling parameters $\Xi\rightarrow \infty$ \cite{Naji,hoda,Netz}. While some aspects of the SC theory for cylindrical macromolecules were studied in previous works \cite{NajiNetzEPJE,Naji_CCT,arnoldholm,NajiArnold}, other aspects have remained unadressed. Our work is aimed specifically at addressing the effects due to image charges, 
which indeed turn out to be quite significant in the SC limit. Using a generalized SC theory and extensive MC simulations we have shown that the
counterions  are strongly depleted away from the cylindrical cores due to repulsion from image charges leading to stronger effective 
repulsions between the two cylinders at smaller separations. The counterion-mediated attraction will be present at intermediate to large separation when
the coupling parameter is sufficiently large. These features are in marked contrast with those obtained with the WC theory. We have mapped a global phase
diagram where attractive and repulsive interactions emerge between two like-charged cylinders with or without the image-charge effects. 

Here we have employed a first-order image approximation to deal with the image-charge effects within the SC theory (note, however, that the numerical scheme used within 
the WC limit allows to account for the full effect of the dielectric discontinuity). 
Generally, the treatment of the image charges in nontrivial geometries becomes very complicated and the two-cylinder model is no exception. The first-order image 
approximation (\ref{Vim2}) has the obvious advantage that it allows  for an analytical treatment of the system. The study of the full effect due to the dielectric discontinuity 
in the two-cylinder model requires more advanced numerical and analytical developments that might become available in the future. 
Another interesting effect which can be investigated in the present context is that of the additional salt that may be present besides the neutralizing counterions. 
Recent generalizations \cite{SC-DHMATEJ} should allow for a systematic study of salt screening effects within the WC-SC framework.

\section{Acknowledgement}

R.P. would like to acknowledge the financial support by the Slovenian Research Agency under contract
Nr. P1-0055 (Biophysics of Polymers, Membranes, Gels, Colloids and Cells). 
M.K. would like to acknowledge the financial support by the Slovenian Research Agency under the young researcher grant. 
A.N. is a Newton International Fellow. 

\appendix

\section{Cross interaction term $v_{\rm cross}(\Av r)$}
\label{app:cross}

In this Section, we explain how the  cross energy term (\ref{vcross}) may be evaluated in the case of two charged cylinders (Section \ref{subsec:SC_2cyln}). 
This requires employing some mathematical identities as we shall explain below. 
Note that here the position vectors centered at two different cylinders appear simultaneously, $\mu\ne \nu$,
\begin{equation}
v_{\rm cross}(\Av r_\mu)=\beta e_0 q\int u_{\rm im}(\Av r_\mu,\Av r_\mu')\sigma(\Av r_\nu')\rmd \Av r'.
\label{app:vcross_def}
\end{equation} 
Inserting the image kernel $u_{\rm im}(\Av r, \Av r')$, Eq.~(\ref{Vim}), and the cylinder charge density $\sigma(\Av r)$, Eq.~(\ref{n0}), we first transform the coordinates 
$\rho_\mu$ and $\varphi_\mu$ centered at the $\mu$-th cylinder to the those centered at the $\nu$-th cylinder using the Graf's addition theorems \cite{langbein},
\begin{eqnarray}
K_m(k\rho_\mu)\sin\,m\varphi_\mu&=&\sum_{n=0}^\infty S_{mn}(k\rho_\nu)\sin\,n\varphi_\nu,\\
K_m(k\rho_\mu)\cos\,m\varphi_\mu&=&\sum_{n=0}^\infty C_{mn}(k\rho_\nu)\cos\,n\varphi_\nu,
\end{eqnarray}
with $m$ and $n$ being integers and 
\begin{eqnarray}
S_{mn}(k\rho_\nu)&=&[K_{m+n}(kR)-K_{m-n}(kR)]I_n(k\rho_\nu)\hspace{3ex}\\
C_{mn}(k\rho_\nu)&=&[K_{m+n}(kR)+K_{m-n}(kR)]I_n(k\rho_\nu)\nonumber\\
	&&\times\> (1-\frac{1}{2}\delta_{n0}).
\end{eqnarray}
We can then carry out the integration in (\ref{app:vcross_def}) over $\rho_\mu'$, $\varphi_\mu'$ and $z'$. Integration over $z'$ produces a $(2\sin k z_\infty)/k$ term, where $z_\infty\to\infty$ is the upper limit and thus we obtain
\begin{eqnarray}
v_{\rm cross}(\Av r_\mu)&=&
\frac {4\xi}\pi \sum_{m=0}^\infty \int_0^\infty\rmd k\,\xi_m(ka)K_m(k\rho_\mu)\nonumber\\
	&&\times\>\frac{\sin k z_\infty}{k}\,\,C_{m0}(ka)\cos m\varphi_\mu. 
\end{eqnarray}
In the next step we integrate over the wave-vector $k$, where the integrand contains the rapidly oscillating factor $\sin k z_\infty$, which supresses contributions of the integrand in whole the range, except at $k\to 0$. Formally, we use the following mathematical identity 
\begin{equation}
\lim_{\omega_0\to \infty}\int_0^\infty f(t)\sin\,\omega_0 t\,\rmd t=\frac{\pi}{2}\,\textrm{Res}(f,0),
\end{equation}
where $\textrm{Res}(f,0)$ is the residue of function $f(t)$ at $t=0$. This integral is finite only if $f(t)$ behaves as $1/t$ for  $t\to 0$ and goes to 0 for $t\to\infty$.
This statement can be proven by considering the split of function $f(t)$ into a $1/t$ term and a remainder, i.e., $f(t)=t^{-1}\textrm{Res}(f,0)+\tilde f(t)$. The first part can be integrated straightforwardly, whearas the remainder yields 0 according to the Riemann-Lebesgue lemma \cite{springer}. This finally leads us to 
\begin{equation}
\label{app:vcross}
v_{\rm cross}(\Av r_\mu) = 2\Delta\xi\,\sum_{m=1}^\infty \frac{1}{m}\Bigl(\frac{a^2}{R\rho_\mu}\Bigr)^m\cos\, m\varphi_\mu,
\end{equation}
which may be simplified further using the identity 
\begin{eqnarray}
&& 2 \sum_{m=1}^\infty \frac{1}{m}\Bigl(\frac{a^2}{R\rho_\mu}\Bigr)^m\cos\, m\varphi_\mu=
\label{app:vcross_2}\\
&&\qquad \qquad  = - \ln\Bigl[1-2\Bigl(\frac{a^2}{R\rho_\mu}\Bigr)\cos\,\varphi_\mu+ \Bigl(\frac{a^2}{R\rho_\mu}\Bigr)^2\Bigr]\nonumber
\end{eqnarray}
and yields  the result given in  Eq.~(\ref{vcross}).

\section{MC simulations}
\label{app:MC}

We performed Monte-Carlo simulations in order to study the system of one and two charged dielectric cylinders at arbitrary coupling parameters, which thus go 
beyond the weak and strong coupling theories. All simultations were performed in the canonical ensemble ($NVT$) using standard Metropolis algorithm \cite{Metropolis:53, MCfrenkel}.
In the case of one charged cylinder the system (including the neutralizing counterions) is enclosed in a cylindrical simulation box of outer radius $a_{\mathrm{out}}=50a$ whereas 
in the case of two cylinders the system is enclosed in a square box of lateral size $L_\bot/a=60$ (WC) and 100 (SC)  
(note that in this work we focus on the regime of large Manning parameters where
the lateral size and shape of the confining box becomes unimportant  \cite{NajiNetzEPJE,Naji_CCT}). The simulation box height is assumed to have a finite value $L_z$, which in terms of other
physical parameters by invoking the global electroneutrality condition: 
in the case of a single cylinder it is given by $L_z = Nqe_0/\lambda$, and in the case of two cylinders $L_z = Nqe_0/(2\lambda)$. 
We use periodic boundary conditions in the $z$ direction by replicating the main simulation box
infinitely many times in that direction. 

The energy of the system for a given configuration of the cylinder(s)  is composed of two main parts
\begin{equation}
\beta W=\sum_{i=1}^N \beta W_{0c}(\Av r_i)+\sum_{i\ge 1}^N\beta W_{cc}(\Av r_i,\Av r_j)
\label{total_en}
\end{equation}
The first term is the counterion-cylinder interaction energy as given by Eqs.~(\ref{eq_W0c}) and (\ref{eq:W0c_two}) in the text.
The second term in (\ref{total_en}) is counterion-counterion interaction energy which includes also the contribution from interactions with counterion 
image charges. Due to the  periodic boundary conditions used in the simulations, this latter term involves infinite summation series \cite{mmm2d}. This is because
the counterions and image charges in the main simulation box interact also with their  periodic ``copies" 
as obtained by the replication of the main simulation box to an infinite number of simulation box copies. These summations can be evaluated as explained in the 
forthcoming Sections. 

The interaction energy  between two given counterions $i$ and $j$ can be written as 
\begin{equation}
\beta W_{cc}(\Av r_i,\Av r_j)=
	\left\{
	\begin{array}{ll}
        \infty&\quad\!\!\!\! \vert\Av r_i-\Av r_j\vert\le 2R_c,\\
	&\\
	\frac 12 w_{\rm im}(\Av r_i,\Av r_i)&\quad\!\!\!\! i=j,\\
	&\\
	w_0(\Av r_i,\Av r_j)+w_{\rm im}(\Av r_i,\Av r_j)&\quad\!\!\!\!\textrm{otherwise}.\\
	\end{array}\right.
\label{ionion}
\end{equation}
The first equation corresponds to the hard-core repulsion between two counterions overlapping counterions. 
The second equation gives the image self-energy of the $i$-th counterion, i.e., it 
takes into account the interaction of a given counterion with its dielectrically induced images 
in the polarizable dielectric core. 
The third equation in (\ref{ionion}) gives the interaction between two different counterions $i$ and $j$. It is written as the sum of two
different contributions which will be calculated below: i) the Coulomb interaction, $w_0$, between the $i$-th (located at $\Av r_i$) and the $j$-th counterions and
all its periodic copies (located at $\Av r_j + k L_z\,\Av e_z$ for integer $k=-\infty,\ldots,\infty$),
\begin{equation}
w_0(\Av r_i,\Av r_j)=\sum_{k=-\infty}^\infty\frac{q^2 \ell_{\mathrm{B}}}{\vert\Av r_i-\Av r_j+k L_z\,\Av e_z\vert}
\end{equation}
and ii) the contribution involving interactions with the dielectrically induced image charges $w_{\rm im}$ that will be defined later in this Appendix. 

\subsection{Calculation of $w_0$}

The first contribution $w_0$ can be written as 
\begin{equation}
w_0(\Av r_i,\Av r_j)
	 = \left\{	\begin{array}{ll}
	w_D(\Av r_i,\Av r_j),	&\quad \Delta\rho/L_z<\Delta\rho_0,\\
	w_L(\Av r_i,\Av r_j),	&\quad \Delta\rho/L_z>\Delta\rho_0,\\
	\end{array}
\right.
\end{equation}
where we have introduced  two different summation schemes 
depending on the radial separation $\Delta\rho$ between the two position vectors $\Av r_i$ and $\Av r_j$ 
in order to increase the convergence of the sum over the long-range electrostatic interactions.
If radial distance $\Delta\rho=\sqrt{\Delta x^2+\Delta y^2}$ is smaller than a threshold $\Delta\rho_0$, 
we first sum up $k_0$ terms directly and then use an approximation to estimate the remaining part of the series (from $k_0+1$ to infinity). We refer to this as the ``direct" summation scheme. This method is different from the so-called Sperb \cite{sperb} summation scheme which
involves calculating various special functions. For our purposes of relative precision up to $10^{-5}$ we found it more efficient and can be made in principle as accurate as necessary. 
It follows as 
\begin{equation}
\frac{w_D(\Av r_i,\Av r_j)}{q^2 \ell_{\mathrm{B}}}=\sum_{k=-k_0}^{k_0}\!\! s(k)+\frac{2\Delta z^2-\Delta\rho^2}{L_z^3}\sum_{k=k_0+1}^{\infty}\frac{1}{k^3},
\end{equation}
where the last sum over $1/k^3$ is equal to $-\frac{1}{2}\psi''(k_0+1)$, where $\psi$ is digamma function, and needs to be calculated
only once in the course of the simulations. The term $s(k)$ is defined as 
\begin{equation}
s(k)=\frac{1}{\sqrt{\Delta\rho^2+(\Delta z+k L_z)^2}}.
\end{equation}
This scheme 
gives an error of the order $O(1/k_0^4)$.

For larger radial separations ($\Delta\rho>\Delta\rho_0$), we use the so-called Lekner summation scheme \cite{lekner}
as it converges more rapidly. It is given by 
\begin{eqnarray}
\frac{w_L}{q^2 \ell_{\mathrm{B}}}&=& -\frac{2}{L_z}\ln\,\frac{\Delta\rho}{L_z}+\frac{4}{L_z}\sum_{m=1}^{\infty}K_0(k_m \Delta\rho)\cos(k_m\Delta z)+\nonumber\\
&&+\Delta w_{\rm offset},
\end{eqnarray}
where $\Delta w_{\rm offset}$ accounts for the  offset generated at the threshold, i.e., 
\begin{equation}
\Delta w_{\rm offset}=w_D(\Delta\rho=\Delta\rho_0)-w_L(\Delta\rho=\Delta\rho_0). 
\end{equation}

We choose the threshold value as $\Delta \rho_{\perp 0}=0.4 L_z$ and use a cut off of $k_0=6$ in the direct scheme  
and sum up to $2+[1.5/\Delta \rho_\perp]$ terms in the Lekner scheme, which give  a relative error 
smaller than $10^{-5}$. 


\subsection{Calculation of $w_{\rm im}$}
\label{app:images}

Let us now consider the second contribution that enters Eq.~(\ref{ionion}), i.e., $w_{\rm im}(\Av r_i,\Av r_j)$, which is the interaction 
energy obtained by summing up all contributions that involve dielectrically induced image charges throughout different periodic copies
of the main simulation box along the $z$ axis.  
It can be written in the case of one cylinder as 
\begin{equation}
w_{\rm im}(\Av r_i,\Av r_j)=\beta(e_0 q)^2\!\!\sum_{n=-\infty}^\infty u_{\rm im}(\Av r_i,\Av r_j+nL_z{\Av e}_z),
\label{w_im_def}
\end{equation}
where ${\Av e}_z$ is a unit vector pointing in $+z$ direction.
Inserting the image kernel (\ref{Vim}) into the above equation, one ends up with an infinite summation of terms including $\cos\,k(\Delta z+nL_z)$, which can be expressed in terms of $\delta$ functions, i.e., 
\begin{equation}
\sum_{n=-\infty}^\infty\cos\,k(\Delta z+n L_z)=\pi\cos\,k\Delta z\sum_{n=-\infty}^\infty\delta(kL_z-2\pi n).
\end{equation}
This enables us to simply integrate over the wave-vector variable  $k$ and thus obtain 
\begin{eqnarray}
\frac{w_{\rm im}(\Av r_i, \Av r_j)}{q^2 \ell_{\mathrm{B}}}&=&
	\frac{2}{L_z}\sum_{m=0}^\infty\sum_{n=-\infty}^\infty\xi_{m}(k_n a)K_m(k_n\rho_i)K_m(k_n\rho_j)\nonumber\\
		&&\times\>\cos\, k_n\Delta z\, \cos\, m\Delta \varphi,
\end{eqnarray}
where $k_n=2\pi n/L_z$.
The $n=0$ term needs to be handled carefully by taking the limit $k_n\to 0$ as
\begin{equation}
\lim_{k\to 0}\xi_{m}(k_n a)K_m(k_n\rho_i)K_m(k_n\rho_j)=\frac{\Delta}{m}\Bigl(\frac{a^2}{\rho_i\rho_j}\Bigr)^m.
\end{equation}
Using the summation identity as used in Eq.~(\ref{app:vcross_2}), we 
find the final expression 
\begin{eqnarray}
\frac{w_{\rm im}(\Av r_i, \Av r_j)}{q^2 \ell_{\mathrm{B}}}&=&-\frac{\Delta}{L_z}\, \ln\left[1-2\Bigl(\frac{a^2}{\rho_i\rho_j}\Bigr)\cos\Delta\varphi+\Bigl(\frac{a^2}{\rho_i\rho_j}\Bigr)^2\right]+\nonumber\\
&&+\frac{4}{L_z}\sum_{m=0}^\infty\sum_{n=1}^\infty\xi_{m}(k_n a)K_m(k_n\rho_i)K_m(k_n\rho_j)\nonumber\\
&&\times\>\cos\, k_n\Delta z\, \cos\, m\Delta \varphi.
\label{app:imagesum}
\end{eqnarray}
This expression is the most computationally expensive part of the simulations, and its efficiency decreases for $\rho_i/L_z\ll 1$, i.e., for counterions near the cylindrical core.
In the simulations we truncate these summations in such a way that the relative cut-off error is less than $10^{-4}$.

In the case of two cylinders, we use the first-order image approximation in the simulations as well, 
and thus the image-charge contributions are simply the sum of the contributions for each cylinder separately, i.e., 
\begin{equation}
w^{(2)}_{\rm im}(\Av r_i,\Av r_j)=w_{\rm im}(\Av r_{i,1},\Av r_{j,1})+w_{\rm im}(\Av r_{i,2},\Av r_{j,2}),
\end{equation}
where indices 1 and 2 represent coordinates centered at the first and the second cylinder, respectively.

\section{Calculation of the inter-cylinder force in the simulations}
\label{app:MC_force}

In this Section, we explain the method used to directly evaluate the force between two charged cylinders within MC simulations. 
The force acting on a given  cylinder (say cylinder 1) is composed of three contributions, 
\begin{equation}
F=F_{00}+\sum_{i=1}^N F_{0c}(i)+\sum_{i=1}^N F_{\rm osm}(i),
\label{force}
\end{equation}
where $F_{00}$ is the bare electrostatic force due to the interaction with the other cylinder, 
$F_{0c}(i)$, is the electrostatic force due to the interaction of the cylinder with the $i$-th counterion (summed over all counterions $i=1,\ldots,N$), 
and $F_{\rm osm}(i)$ is the force due to the osmotic pressure from the $i$-th counterion. 

The first contribution follows from the interaction energy $W_{00}$,  Eq.~(\ref{W00_delta}),  as 
\begin{equation}
F_{00}=-\frac{\partial W_{00}}{\partial R},
\end{equation} 
which gives
\begin{equation}
\frac{\beta F_{00}}{L}=\frac{2\xi^2}{q^2\ell_{\mathrm{B}} R}\Bigl(1+\frac{2\Delta a^2}{R^2-a^2}\Bigr). 
\label{F00}
\end{equation}
The second contribution follows by differentiating $W_{0c}$, Eq.~(\ref{eq:W0c_two}), with respect to $R$ by assuming that the cylinder 1 and
the $I$-th counterion are fixed, i.e., 
\begin{equation}
F_{0c}(i)=-\left.\frac{\partial W_{0c}}{\partial R}\right\vert_{\rho_{2,i},\rho_{2,i}^*,\varphi_{2,i},\varphi_{2,i}^*=\textrm{const}}
\end{equation}
which after some algebra gives
\begin{eqnarray}
\frac{\beta F_{0c}(i)}{L}&=& -\frac{2\xi}{L_z}\biggl[(1-\Delta)\frac{\cos\varphi_{i,1}}{\rho_{i,1}}\nonumber\\ &&+\Delta\Bigl(1+\frac{a^2}{R^2}\Bigr)\frac{\cos\varphi_{i,1}^*}{\rho_{i,1}^*}
+\Delta \frac{a^2}{R^2}\frac{\cos\varphi_{i,2}^*}{\rho_{i,2}^*}\biggr].\qquad
\end{eqnarray}
Here $\rho_{i,1}$ and $\varphi_{i,1}$ are the radial distance and the azimuthal angle of the $i$-th counterion position with respect to cylinder 1, 
and $\rho^*_{i,1}$ and $\varphi^*_{i,1}$ are the corresponding coordinates with respect to the shifted axis by $a^2/R$ from the axis of cylinder 1 toward
the axis of the cylinder 2, Fig.~\ref{fig:model_2cyln}. 
Similar definitions apply to $\rho_{i,2}$, $\varphi_{i,2}$, $\rho^*_{i,2}$ and $\varphi^*_{i,2}$. 

The osmotic (third) term in Eq.~(\ref{force}) results from the collisions of counterions with the cylinder. This contribution amounts to 
a pressure of counterions exerted on the cylinder surface, which is proportional to the contact density of counterions at the cylinder surface  $n_0$. 
Thus the infinitesimal force acting on the cylinder is given by $\rmd F=n_0(\varphi) kT a\rmd \varphi\,L\cos\varphi$. Note that 
the contact counterion density $n_0(\varphi)$ 
is given by the number of counterions inside a small box of dimensions $a\,\delta\varphi\times\delta \rho\times L$ next to the cylindrical surface.
The total osmotic force of all ions in simulation cell is then
\begin{equation}
\frac{\beta F_{\rm osm}}{L}=\lim_{\delta \rho\to 0}\,\frac 1{L_z}\sum_i\frac{\cos\varphi_{i,1}}{\delta \rho}\,\Theta(a+\delta\rho-\rho_{i,1}), 
\end{equation}
where $\Theta$ is the Heaviside step function which is $1$ only if the $i$-th counterion falls inside a shell thickness $\delta \rho$ from the cylinder surface. 
The above procedure in principle gives the exact value of the osmotic force when  $\delta\rho\to 0$. It also gives a convenient method to calculate the
osmotic contribution within MC simulations by evaluating the force for different values of the shell thickness and estimating 
the limiting value by extrapolation.

\end{document}